\documentclass[
 aip,amsmath,amssymb,
 reprint]{revtex4-1}
\usepackage{epsf}

\usepackage{amssymb}
\usepackage{graphics}
\usepackage{color}
\usepackage{subcaption}
\usepackage{graphicx}
\usepackage{float}

\usepackage{verbatim}

\usepackage{rotating}
\usepackage{dcolumn}
\usepackage{bm}
\usepackage{enumitem}

\graphicspath{{./Figs_eps}}

\usepackage{newtxtext}
\usepackage[colorlinks=true, pdfstartview=FitV, linkcolor=blue,
            citecolor=blue, urlcolor=blue]{hyperref} 

\usepackage[utf8]{inputenc}
\usepackage[T1]{fontenc}
\usepackage{amsthm}
\newtheorem{defi}{Definition}

\usepackage{sidecap}
\usepackage{grffile}
\usepackage{scalerel}

\usepackage{xcolor}

\usepackage[mathlines]{lineno}

\def\bea{\begin{equation} \begin{aligned}}
\def\eea{\end{aligned} \end{equation}}
\def\beas{\begin{equation*} \begin{aligned}}
\def\eeas{\end{aligned} \end{equation*}}
\def\bes{\begin{equation*}}
\def\ees{\end{equation*}}

\def\d{\mathrm{d}}

\def\be{\begin{equation}}
\def\ee{\end{equation}}

\begin{document}

\preprint{AIP/123-QED}

\title{Random templex encodes topological tipping points in noise-driven chaotic dynamics}
\author{Gisela D. Charó}
\affiliation{CONICET – Universidad de Buenos Aires. Centro de Investigaciones
del Mar y la Atmósfera (CIMA), C1428EGA  Ciudad Autónoma de Buenos Aires, Argentina}
\affiliation{CNRS – IRD – CONICET – UBA. Institut Franco-Argentin d'Études sur le Climat et ses Impacts (IRL 3351 IFAECI), C1428EGA Ciudad Autónoma de Buenos Aires, Argentina.}

\author{Michael Ghil}
\affiliation{Geosciences Department and Laboratoire de M\'et\'eorologie Dynamique (CNRS and IPSL), \'Ecole Normale Sup\'erieure and PSL University, 75231 Paris Cedex 05, France.}
\affiliation{Department of Atmospheric \& Oceanic Sciences, University of California, Los Angeles, CA 90095-1565, USA.}

\author{Denisse Sciamarella}
\affiliation{CNRS – Centre National de la Recherche Scientifique, 75016 Paris, France.}%
\affiliation{CNRS – IRD – CONICET – UBA. Institut Franco-Argentin d'Études sur le Climat et ses Impacts (IRL 3351 IFAECI), C1428EGA  CABA, Argentina.}
\email{denisse.sciamarella@cnrs.fr}

\date{\today}
\begin{abstract}
Random attractors are the time-evolving pullback attractors of stochastically perturbed, deterministically chaotic dynamical systems. These attractors have a structure that changes in time, and that has been characterized recently using {\sc BraMAH} cell complexes and their homology groups. This description has been further improved for their deterministic counterparts by endowing the cell complex with a directed graph, which encodes the order in which the cells in the complex are visited by the flow in phase space. A templex is a mathematical object formed by a complex and a digraph; it provides a finer description of deterministically chaotic attractors and permits their accurate classification. In a deterministic framework, the digraph of the templex connects cells within a single complex for all time. Here, we introduce the stochastic version of a templex. In a random templex, there is one complex per snapshot of the random attractor and the digraph connects the generators or ``holes'' of successive cell complexes. Tipping points appear in a random templex as drastic changes of its holes in motion, namely their birth, splitting, merging, or death. This paper introduces and computes the random templex for the noise-driven Lorenz system's random attractor (LORA). 
\end{abstract}

\maketitle

\begin{quotation}
Branched manifolds underlying chaotic attractors have topological properties that remain invariant in a deterministic framework, and that can be characterized using homologies.\cite{birman1983knotted, williams1974expanding} A more complete description is obtained if the cell complex whose homologies are computed is endowed with a directed graph (digraph) that prescribes cell connections in terms of the flow direction. Such a topological description is given by a templex, which carries the information of the structure of the branched manifold, as well as information on the flow. \cite{charo2022templex} This work revisits the templex in a stochastic framework. Stochastic attractors in the pullback approach --- like the LOrenz Random Attractor (LORA) \cite{chekroun2011stochastic} --- include sharp transitions in a Branched Manifold Analysis through Homologies (BraMAH) \cite{sciamarella1999topological, sciamarella2001unveiling,charo2021noise}. These sharp transitions can be suitably described using what we call here a random templex, computed from a sequence of BraMAH cell complexes and a digraph. 
The BraMAH cell complexes are such that changes can be followed in terms of how the generators of the homology groups, the ``holes'' of theses complexes, evolve. 
The nodes of the digraph are the generators of the homology groups, and its directed edges indicate the correspondence between holes from one snapshot to the next. Topological tipping points can be identified with the creation, destruction, splitting or merging of holes, through a definition in terms of the nodes in the digraph. 
\end{quotation}

\section{Introduction}\label{sec:intro}

The topological characterization of noise-driven chaos is a challenging issue that is crucial in the understanding of complex systems, where part of the dynamics remains unresolved and is modeled as noise. While additive noise in a system of equations will blur the topological structure, multiplicative noise may radically change it, as shown by Charó et al  \citep{charo2021noise}. These authors extended the concept of a branched manifold to account for the integer-dimensional set in phase space that robustly supports the system's invariant measure at each instant. 

Such a branched manifold, however, does not contain --- as does its deterministic counterpart --- any information about the future or the past of the invariant measure. In other words, the evolution of the system is not completely described by the branched manifold, which is now itself time-dependent. 
The latter requires, therefore, additional information for its complete description.

The templex was introduced in the realm of deterministic attractors in order to provide more topological information than that contained in a cell complex.\cite{charo2022templex} This missing information concerns the flow around the branched manifold. This information can be spelled out using a digraph \cite{bang2008digraphs} that connects the cells of the complex according to the flow. 

But what about the flow in a cell complex representing the invariant measure of a random attractor? One could try to pose it in terms of connections between cells of different cell complexes, but algebraic topology definitions are such that the number and distribution of cells in a cell complex are arbitrary. It is not the individual cells, 
but the topological properties of each cell complex that characterize the changes from one instant to the next. 
Such topological properties are encoded by the generators of the homology groups of the cell complex, and also by the torsion groups. Herein we concentrate on the homology groups exclusively and leave torsions for future work.  The properties of interest are independent of the particular cell decomposition that is adopted to build the cell complex from data. Homologies hence enable us to connect a cell complex of a random attractor at a certain instant, with a cell complex corresponding to a different instant.

Homological properties can be computed at different times, and can also be tracked across sufficiently close time steps, thus helping us detect sudden changes in the topology. The information of how the holes at a certain instant map on the holes at the next time step can be encoded using a digraph. This naturally leads to the definition of a ``random templex'' as the mathematical object that condenses the topological information regarding the evolution of the system's invariant measure in a finite time window. Topological tipping points (TTPs) are contained in a random templex in the form of creation, splitting, merging and destruction of generators or, equivalently, in certain simple characteristics of the nodes of the digraph associated with a sequence of cell complexes. 

This paper provides the theoretical background that leads to the concept of a random templex and shows how to compute it for the Lorenz Random Attractor (LORA). A brief explanation concerning templexes in the deterministic framework appears in section~\ref{sec2}. The construction of a random templex is discussed in section~\ref{sec3} and illustrated by application to LORA. Finally, in section~\ref*{sectemplex}, using the LORA templex of the previous section, we define rigorously the TTPs introduced in a merely intuitive fashion by \citet{charo2021noise} and compute them for a time window that contains the various types of TTPs. The last section contains conclusions and perspectives, including possible applications to the effects of global change on climatic subsystems, often referred to recently as tipping elements \cite{Lenton.ea.2008}. An appendix includes several technical details of the algebraic calculations involved in determining the homologies and the flow on a random templex.

\section{The deterministic framework}
\label{sec2}

Homology theory allows us to classify manifolds in terms of a small number of topological properties \cite{Poincare1895}. In order to study the topological structure of solutions of deterministic dynamical systems, the characterization must be done from a set of discretely sampled trajectories, i.e. a set of points with as many coordinates as time-dependent variables. In the case of the Lorenz \cite{lorenz1963lorenz} chaotic attractor, for instance, the points have three coordinates; we refer hereafter to this model as L63. The phase space is three dimensional, but the points lie on a  butterfly-shaped surface. How can the topological properties of this surface be computed from the set of points? This is achieved building a cell complex. 

\subsection{Cell complexes}
\label{ssec:complex}

A cell complex $K$ is a layered structure formed by a set of $k$-cells with  $k=0, 1, ..., d$. Each cell stands for a Euclidean closed set with a certain dimension: points are $0$-cells, segments are $1$-cells, filled polygons are $2$-cells and so forth. The highest cell dimension defines the dimension of the complex. 

\begin{figure}
\includegraphics[width=\linewidth]{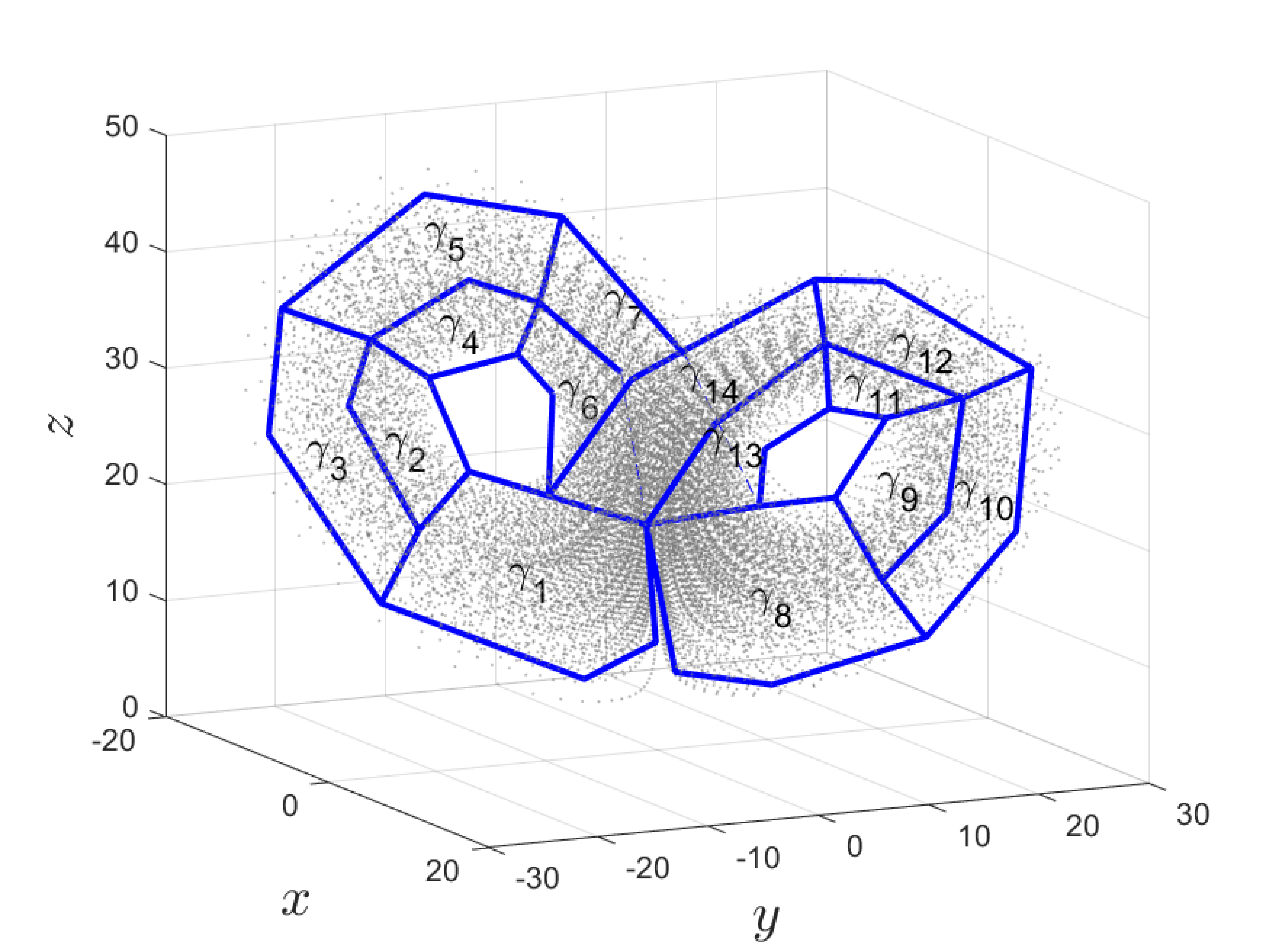}
\caption{Construction of a BraMAH complex for the Lorenz attractor\cite{lorenz1963lorenz}. 
	The point cloud is decomposed into clusters of points that are used to form the 2-complex $K$, 
	with the 2-cells $\{\gamma_i: i=1,...,14\}$. 
}
\label{fig:det_lor}
\end{figure}

There exist methods to build a complex from a point cloud that are quite different. Our choice will be to build a BraMAH complex, i.e. a cell complex whose cells are formed gathering subsets of points which can be locally approximated by a $d$-disk, with $d$ the local dimension of the underlying manifold. Further details of this procedure can be found in \citet{sciamarella2001unveiling}. A BraMAH complex for the Lorenz attractor is shown in Fig.~\ref{fig:det_lor}. Its highest dimensional cell is a $2$-cell and therefore $d=2$. The polygon figures forming the $2$-cells `pave' the surface of the attractor. 
The number of cells used to pave this surface is not fixed. It depends on the criteria used to define the sets of points. But this number has no special significance, because the homologies finally obtained will not depend on the particular size or distribution of the cells \cite{kinsey2012topology}.

In order to make homologies computable, $k$-cells with $k>1$ must have a `direction' or `orientation'. In a $1$-cell this direction can be denoted by an arrow. For instance, if $<1,2>$ stands for instructions to travel along the path from the $0$-cell $<1>$ to the $0$-cell $<2>$, the expression $-<1,2>$ has a natural meaning: it means traveling in the opposite direction, i.e. from  $<2>$ to  $<1>$. A direction or orientation
can also be assigned to $2$-cells: clockwise or counterclockwise. 
A $2$-complex $K$ is said to be directed if each $1$-cell and $2$-cell is assigned a direction. 

A directed complex leads to an algebra of chains. A $k$-chain is a formal linear combination of the $k$-cells in a cell complex. The boundary of a directed $2$-cell, for instance, is the chain formed by the $1$-cells on its boundary, with a positive sign if the direction of the edge is consistent with the direction of the $2$-cell, and with a negative sign otherwise. It is, thus, the linear combination of $1$-cells forming its boundary, while using the signs in accordance with the arrows. All $k$-chains in a complex form an abelian group \cite{kinsey2012topology}.

The purpose of introducing these algebraic concepts is to come up with something that will distinguish the important characteristics of a given cell complex $K$. The chains of $1$-cells forming loops that are not the boundaries of the $2$-cells will be important, as well as a chain of $2$-cells enclosing a cavity, as in a torus or a sphere. These features are summarized by the homology groups of $K$, whose generators identify the `holes' at level $k$ of the cell complex, namely ${\cal H}_k = \left[ \displaystyle h^k_1, ..., h^k_{\beta_k} \right]$, where $h^k_i$ is the $i$th $k$-hole, and $\beta_k$ is the Betti number, which counts the number of $k$-holes. It was Poincar\'e who proved the invariance of these numbers for a given set, independently of the details of the construction of a cell complex for the set. Moreover, $\beta_k = 0$ for $k \ge n$, where $n$ is the dimension of the space in which the set is embedded. \cite{Poincare1895, siersma2012poincare} 

The significance of such homology groups can be understood level by level. At level $0$, we have ${\cal H}_0(K)$, which contains the $0$-holes that  identify the disconnected pieces of the cell complex $K$. L63 has a single connected component and therefore $\beta_0 =1$.

\begin{figure*}
\centering
\begin{subfigure}[b]{0.45\textwidth}
\includegraphics[width=\textwidth]{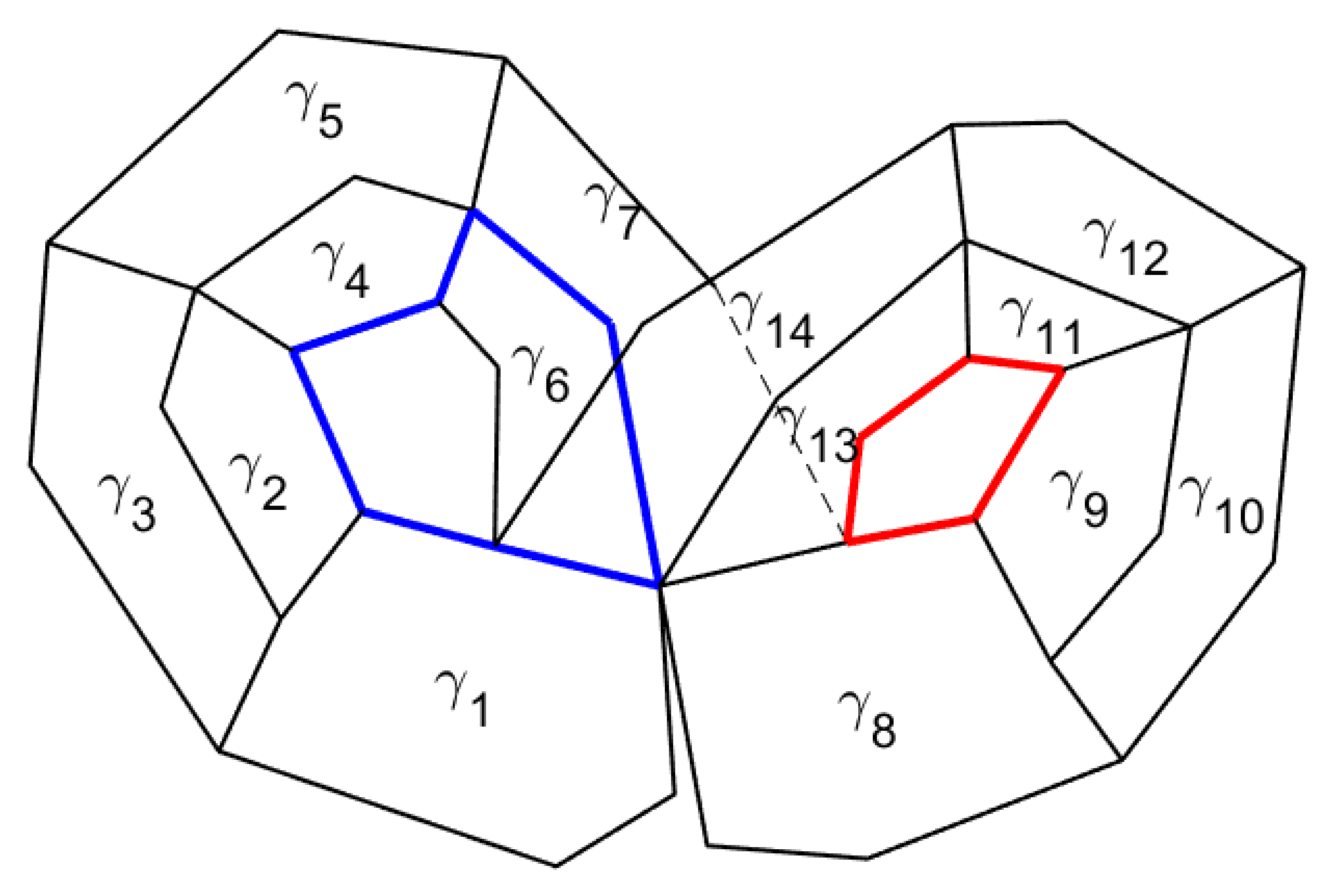}
\caption*{(a)}
\end{subfigure} 
  \begin{subfigure}[b]{0.45\textwidth}
 \includegraphics[width=\textwidth]{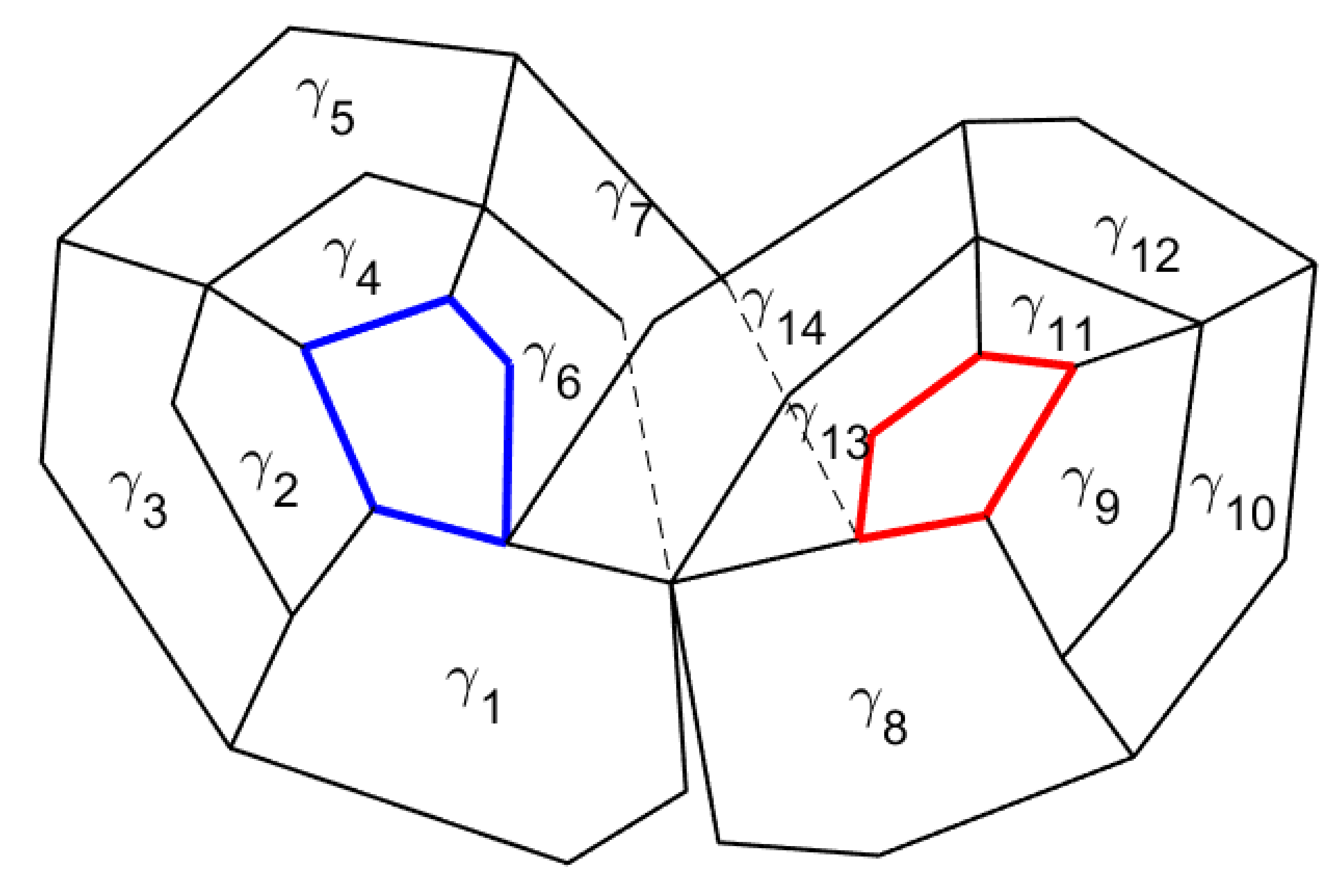}
 \caption*{(b)}
 	\end{subfigure}
\caption{Complex $K$ of Fig.~\ref{fig:det_lor} showing in color 
    (a) the generators of $H_1(K)$ or 1-holes obtained in 
    the homology computation and (b)  the minimal 1-holes.}
\label{fig:holes}
\end{figure*}

Stepping up to $k=1$, ${\cal H}_1(K)$ lists the 1-generators, which are associated with the non-trivial loops or $1$-holes in the complex. When the manifold underlying a complex has ``handles,'' the $1$-generators encircle these handles forming closed sequences of $1$-cells that can be traveled through sequentially. In L63 there are two such handles, $\beta_1 =2$, one in each wing of the butterfly, as shown in Fig.~\ref{fig:holes}. 

These generators need not strictly contour the boundaries of the geometrical hole. A generator may wander around a hole, without tightly encircling the empty space. But the tight holes encircling handles can be retrieved  algebraically, from the 2-complex itself, as long as it is uniformly oriented, i.e., the orientation of the {2-}cells is propagated, so that shared borders (1-cells) of two adjacent 2-cells are canceled out when the borders of all the 2-cells of the complex $K$ are summed. 

This information is contained in what we call the orientability chain ($\mathcal{O}_1$). It is defined as the 1-chain obtained by applying the border operator $\partial_2$ to the sum of all the $2$-cells \cite{sciamarella2001unveiling} (see \ref{homologies}). This yields a sum of $1$-cells, which includes boundaries and torsions, 
\[ \mathcal{O}_1 = \partial_2\left(\sum_i \gamma_i \right) = \sum_j a_j \sigma_j ,\] 
\noindent
where $ \gamma_i$ denotes the 2-cells, $ \sigma_j$ the 1-cells and the $a_j$ are integers. The 1-cells in $\mathcal{O}_1$ whose $a_j = \pm 1$ are the boundaries of the complex. In the case we are dealing with, $\mathcal{O}_1$ can be used to replace the generators by their homologically equivalent ``tight'' holes. These 1-holes will be called minimal holes.

A graphic example comparing generators and minimal holes for the BraMAH complex of the L63 attractor is given in Figs.~\ref{fig:holes}(a) and \ref{fig:holes}(b).   
Comparison of the two panels shows that the 1-generator in the right wing is minimal, while the 1-generator in the left wing is not.

Passing on to the next layer, generators of $H_2$ identify empty volumes enclosed by a surface, here called $2$-holes. This is achieved by looking for chains of $2$-cells forming the set of polygons that enclose the cavity. As there are no cavities enclosed by polygons in the L63 attractor, $\beta_2 =0$, i.e. there are no $2$-holes.  

\subsection{Digraphs and templexes}
\label{ssec:templex}

What does a cell complex lack in order to fully characterize an attractor in phase space? The cell complex describes the shape of the attractor, but not the dynamics of the flow on the attractor. This missing information can be brought into the description by indicating the order in which the cells are visited by the flow. This can be done using a directed graph defined so that its nodes represent the highest dimensional cells of the complex -- the $2$-cells in the case of a $2$-complex. 

\begin{figure}
	\includegraphics[width=\linewidth]{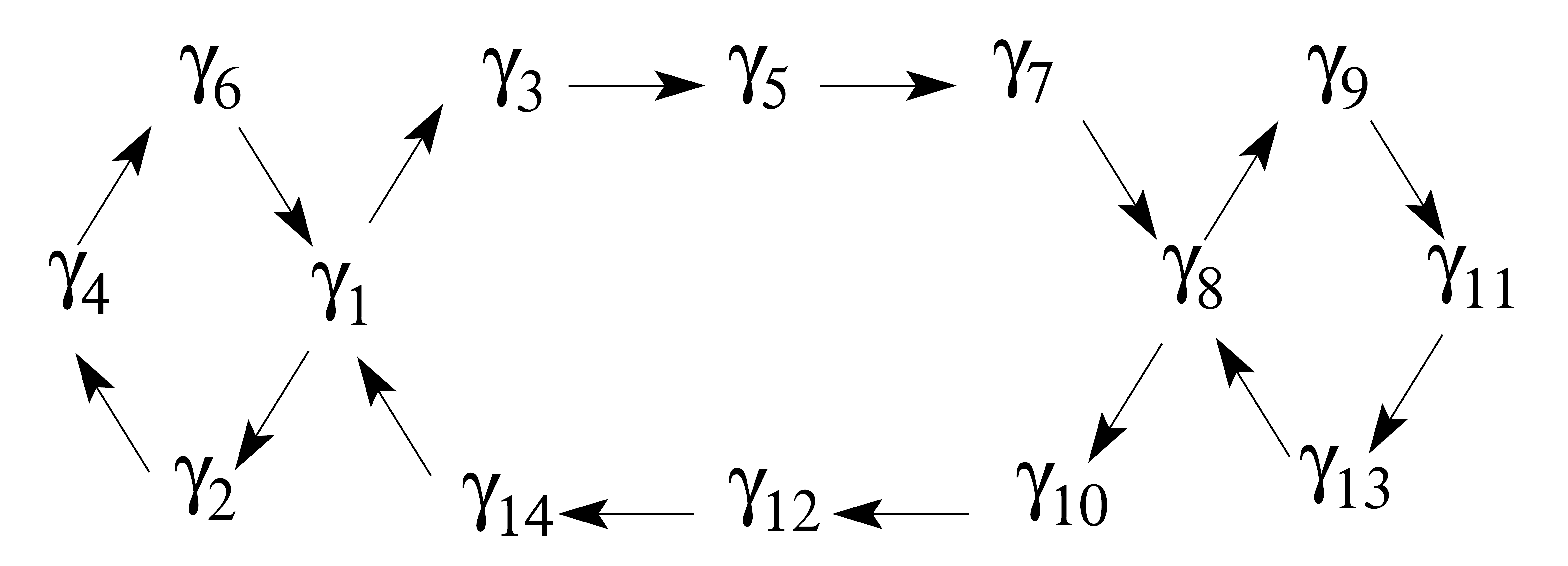}
	\caption{The digraph $G$ for the complex $K$ of L63 shown in Fig.~\ref{fig:holes}. The digraph
		 indicates the connections associated with the flow between the cells. Together, the complex 
		 and the digraph form the templex $T=(K,G)$. 
	}
	\label{fig:digraph_L63}
\end{figure}

Two nodes are linked by a directed edge if the flow connects the cells in a given order. 
In the directed graph for L63 shown in Fig.~\ref{fig:digraph_L63}, there is an edge connecting $\gamma_7$ to $\gamma_{8}$ and $\gamma_{13}$ to $\gamma_8$ because trajectories can flow from the $2$-cells $\gamma_{7}$ and $\gamma_{13}$  to the $2$-cell $\gamma_8$.

This leads to the definition of a dual object formed by a cell complex $K$ and its digraph $G$. Char\'o et al.  \citep{charo2022templex} introduced this novel type of object and called it a {\bf templex}, a contraction of template + complex.  

\begin{defi}
A {\bf 2-templex} ${T} \equiv (K,G)$, is a templex of dimension 2, where $K$ is a 2-complex, whose underlying structure is a branched 2-manifold associated with a {\bf deterministic} dynamical system, and  $G=(N,E)$ is a digraph , such that (i) the nodes $N$ are the 2-cells of $K$ and (ii) the edges $E$ are the connections between the 2-cells associated with  the flow. 
\end{defi}

The properties of a templex that characterize both the shape of the branched manifold and the flow upon it can be derived using the combined properties of the cell complex and the digraph. The result is expressed in terms of what we call {\bf stripexes}. They have the same role and inherit their name from strips in templates \cite{Let95a}. Stripexes identify the nonequivalent ways of circulating around the attractor according to the flow on it. The L63 attractor has four stripexes, which can be computed looking for the cycles in the digraph and retaining only the nonredundant ones. For further details on how to do these computations, the reader is referred to \citet{charo2022templex}. 

The nonequivalent paths along the complex representing the L63 attractor are: 
\begin{subequations}\label{eq:stripex}
	\begin{align}
& \gamma_1 \rightarrow \gamma_2 \rightarrow \gamma_4 \rightarrow \gamma_6 
  \rightarrow \gamma_1, \label{stripex1} \\
& \gamma_8 \rightarrow \gamma_9 \rightarrow \gamma_{11} \rightarrow \gamma_{13}    
\rightarrow \gamma_8,  \label{stripex2} \\
& \gamma_1 \rightarrow \gamma_3 \rightarrow \gamma_5 \rightarrow \gamma_7  
\rightarrow \gamma_8, \label{stripex3} \\
& \gamma_{8} \rightarrow \gamma_{10} \rightarrow \gamma_{12}   \rightarrow \gamma_{14} 
\rightarrow \gamma_{1}, \label{stripex4}
  	\end{align}
\end{subequations}

\noindent
The stripexes of $T=(K,G)$ are the four sub-templexes associated with the distinct paths apparent in Fig.~\ref{fig:digraph_L63}. They are denoted by $\mathcal{S}_i$ ($i=1,2,3,4$). The first two stripexes are paths along either wing of the butterfly, according  to Eqs.~\eqref{stripex1} and \eqref{stripex4}. The last two are complementary paths going from one wing to the other, according  to Eqs.~\eqref{stripex2} and \eqref{stripex3}.  

In a deterministic framework, the topological structure of an attractor can be accurately described by a templex because the branched manifold is invariant. But this is not the case for a random attractor. As shown by \citet{charo2021noise}, the concept of branched manifold and cell complex can still be of help in the stochastic context. The next section will discuss how templex theory can be extended to deal with a time-varying topological structure.

\section{The stochastic framework} \label{sec3}

In a stochastic framework, ensembles of trajectories driven by the same noise path $\omega_t$ can be tracked and provide considerable insight on the overall behavior of the system, as shown by M.D. Chekroun, M. Ghil and E. Simonnet \cite{chekroun2011stochastic,Ghil.ea.2008} and by T. T\'el and associates \cite{Bodai.Tel.2012, tel2020theory} in the climate sciences and by a vast literature on random dynamical systems \cite{crauel1994attractors,arnold1998random}. This pullback approach cancels out the well-known smoothing effect of noise and makes the fractal structure of the noise-driven chaotic dynamics emerge. 

When the L63 model \cite{lorenz1963lorenz} is perturbed by a multiplicative noise in the Itô sense \cite{arnold1998random}, with $W_t$ a Wiener process and $\sigma > 0$ the noise intensity, we get the stochastic Lorenz model of \citet{chekroun2011stochastic}:
\bea\label{eq_slm}
\vspace{-1ex}
&\d x=r(y-x)\d t+\sigma x {\mathrm d} W_t, \\
&\d y=(rx-y-xz)\d t+\sigma y {\mathrm d} W_t, \\
&\d z=(-bz+xy)\d t+\sigma z {\mathrm d} W_t; 
\vspace{-1ex}
\eea
The three parameters take the standard values for deterministically chaotic behavior $(r, \sigma, b) = (28,10,8/3)$. 

In the pullback approach, one is interested in the time-dependent sample measures $\mu_t$ driven by the same noise realization until time $t$, starting from any compact ensemble of initial data. The point clouds that are obtained in this way show the time-dependent stretching and folding mechanisms caused by the noise-driven nonlinearities. These point clouds evolve, combining the smoothness of the L63 deterministic convection with sudden deformations of the pullback attractor's support. In the stochastic setting, these pullback attractors are called random attractors \cite{crauel1994attractors,arnold1998random} or snapshot attractors \cite{romeiras1990multifractal, tel2020theory}.  

In practice, the convergence of the $\mu_t$’s approximation is observed for a set of $N_0=10^8$ initial points. At each time instant $t$, each point in the random attractor is associated with a value of $\mu_t$ that is obtained by averaging over a volume encompassing that point. In order  to characterize the topology of LORA at each time instant, we select a threshold for approximating the sample measure $\mu_t$ \cite{charo2021noise}.

In this paper, we sieved LORA’s point clouds by applying a threshold in computing the sample measure $\mu_t$, which is computed on a grid of the $(y,z)$-plane. We only retain the points whose projection onto this plane is surrounded  by more than $n=100$ points within a grid box of size $\delta y \times  \delta z =0.0684 \times 0.0587$.

\begin{figure*}
	\includegraphics[width=0.7\linewidth]{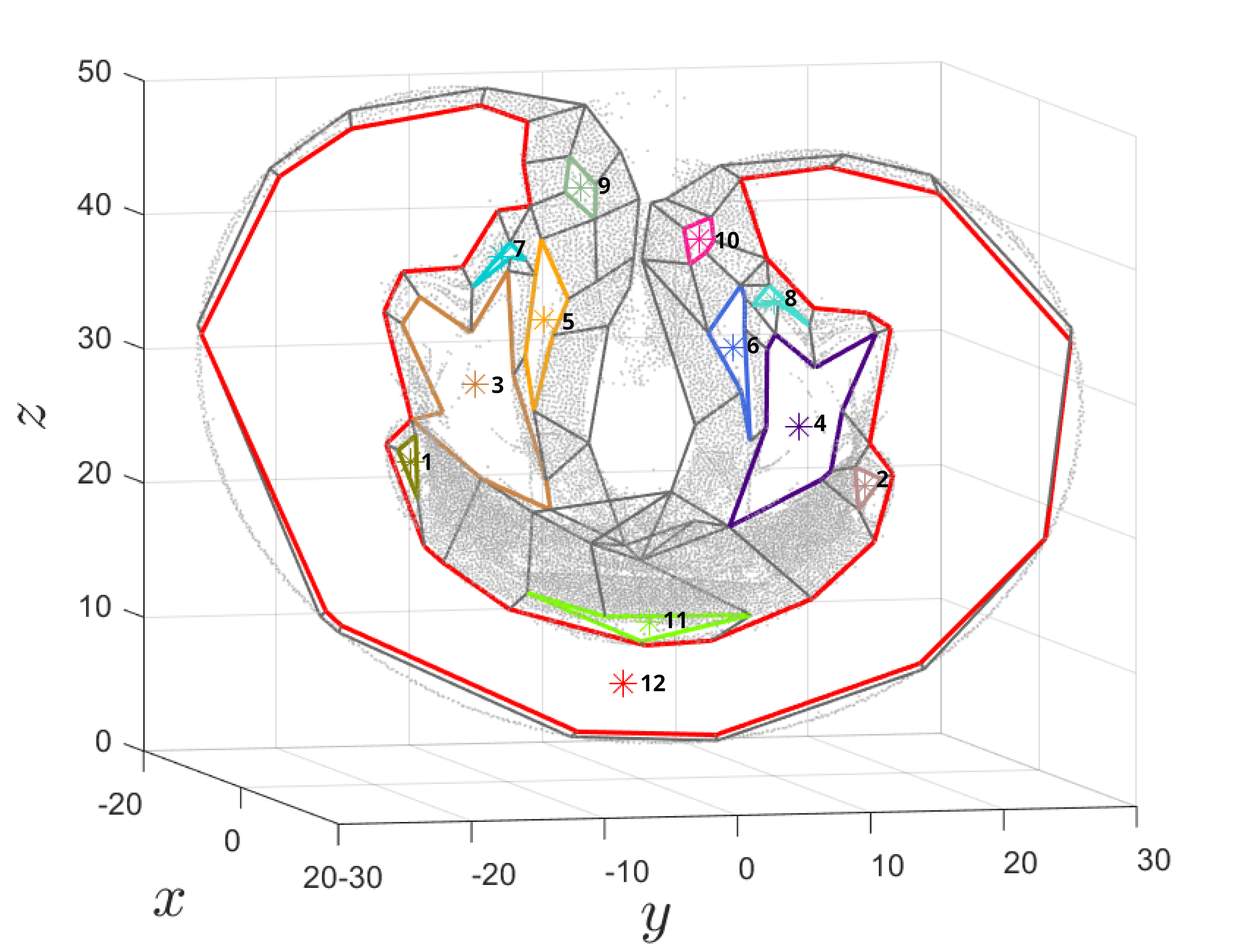}
	\caption{Phase portrait showing the point cloud (in gray) juxtaposed on the cell complex, and the minimal 1-holes (in colors) for a single snapshot of the Lorenz Random Attractor (LORA) 
		at $t=40.09$. There are ten 1-holes and their barycenters are labeled with numbers. 
}
	\label{fig:MinCyc}
\end{figure*}

An example of  minimal holes of a LORA point cloud is provided in Fig.~\ref{fig:MinCyc} for $t=40.09$, $\sigma = 0.3$. For further details on constructing a {\sc BraMAH} complex for a snapshot from a point cloud, see \citet{charo2021noise}.

A new time instant requires the construction of another cell complex. By constructing a few cell complexes and computing their homology groups, LORA was found to undergo abrupt topological changes as it evolves  \cite{charo2021noise}. As already mentioned, we can use cell complexes to show how these topological changes take place, but we can also take a step further in describing these changes. 

Defining a templex for a random attractor involves establishing a link between different snapshots, in order to incorporate time into the description, as done in the deterministic case. But how can the topology of the point cloud at a given instant be related to the topology of another instant? We can compute the approximations $\hat{\mu_t}$ for several time instants, and build cell complexes for each snapshot. Establishing a correspondence between the cells in subsequent cell complexes is not trivial, since, as we have explained in section~\ref{ssec:complex}, the number of cells and their distribution is arbitrary. 

The cell complex as a whole`flows' in phase space for a fixed noise realization $\omega_t$. Due to the fact that the evolution of the deterministic convection is smooth, the 1-holes in the complexes --- which are intrinsic to the snapshot  --- can be followed from one snapshot to the next This can be used to check if the holes are simply displaced or if they are modified drastically, through splitting, merging, creation or destruction events. We are now in a position to propose a definition of a random templex of dimension $2$.

\begin{defi}
A {\bf random 2-templex} $\mathcal{R}$ is an indexed family $\mathcal{K}$ of BraMAH 2-complexes and a digraph $\cal{D}$,  $\mathcal{R} = (\mathcal{K}, \mathcal{D})$, such that:
\begin{itemize}
\item[(i)] The family of $s$ BraMAH 2-complexes corresponds to the approximation of the branched manifold that robustly supports the point clouds associated with the system’s invariant measure $\mu_t$ at each instant t. The family of  point clouds $\{{C_1},... {C_s}\}$ corresponds to the snapshots for the time instants $T_w=[t_1,....,t_s]$.

\item[(ii)] There is one 2-complex $K_j$ per snapshot $t_j$ approximating the branched manifold that underlies the point clouds $\{C_j:  j = 1, \dots, s\}$, with $\mathcal{K} = \{K_j:  j = 1, \dots, s\}$. 

\item[(iii)] In the digraph $\mathcal{D = (N,E)}$ each node in $\mathcal N$ is a minimal hole for a complex $K_j$ in the time window $T_w$, and the edges $\cal{E}$ denote the paths between minimal holes from one $t_j$ to another.   

\end{itemize}
\end{defi}

Notice that the digraph in a random templex does not connect 2-cells in a  single branched manifold, as is the case in the deterministic case of {\bf Definition 1}, but 1-holes between distinct time steps. Holes may just move or deform, so that the geometry of the evolving branched manifold changes without changing the topology. But a hole may disappear from a snapshot to the next or be created, split, or merge. In such cases, homology groups will change and so will the topology of the random attractor. Topological changes 
occur at specific times, which are associated with TTPs, as mentioned in section~\ref{sec:intro}. A random templex will be shown to encode such TTPs. 

We limit our description here to the homology groups through the minimal 1-holes, without discussion of orientability properties, which are associated with the torsion groups of a cell complex. Neglecting torsion properties works here because the LORA complexes do not present distinct torsion groups, which is the case for many other known deterministic attractors, when changing parameters and hence, supposedly, for their corresponding random attractors.
But we do not exclude that torsion groups might be relevant for other random attractors, or that twists as defined in  \citet{charo2022templex} might also play a role.

\begin{figure*}
\begin{subfigure}{0.9\hsize}\centering
    \includegraphics[width=0.9\hsize]{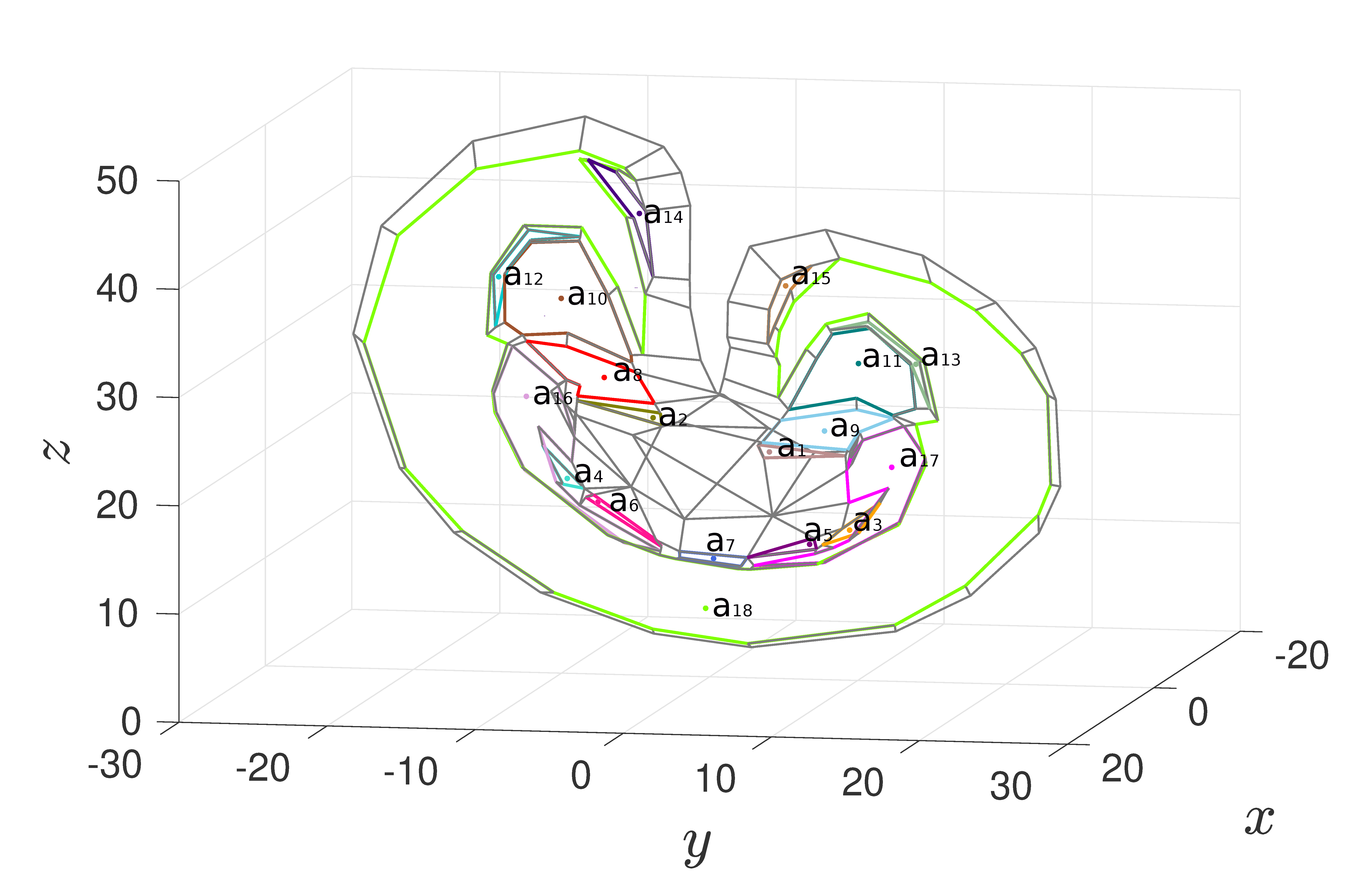}
\caption*{(a)}
    \label{fig:sub1}
\end{subfigure}%

\begin{subfigure}{0.9\hsize}\centering
    \includegraphics[width=0.9\hsize]{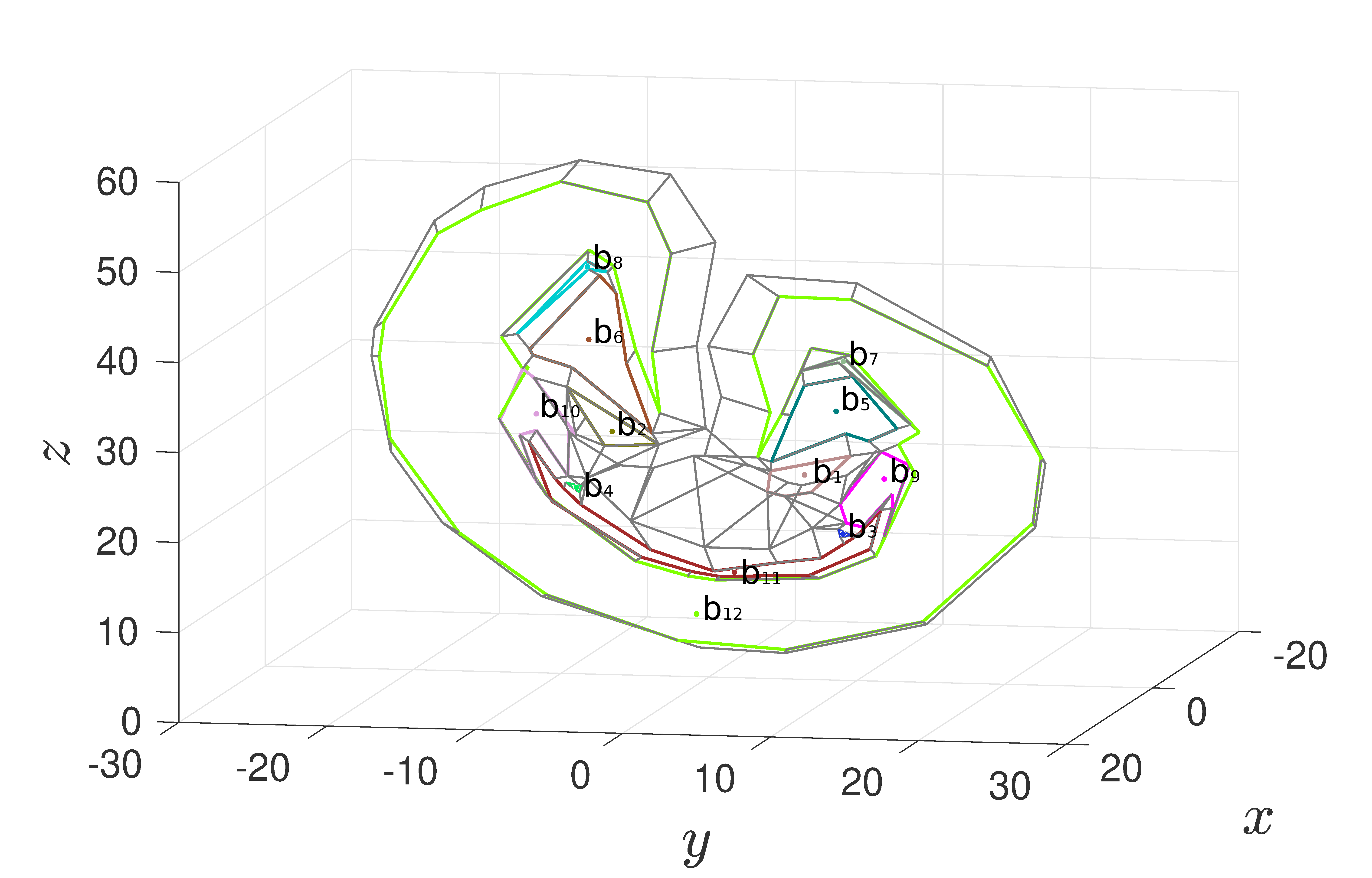}
\caption*{(b)}
    \label{fig:sub2}
\end{subfigure}

\begin{subfigure}{0.5\hsize}\centering
    \includegraphics[width=0.9\hsize]{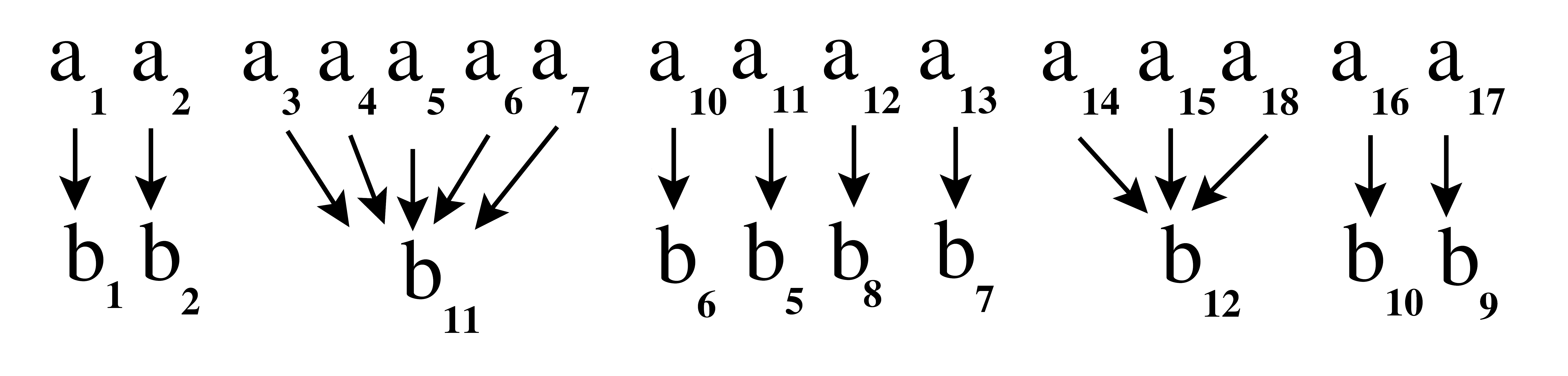}
    \caption*{(c)}
    \label{fig:sub3}
\end{subfigure}

\caption{Minimal 1-hole tracking from a snapshot (a) at $t=40.07$ to the next one (b) at $t=40.075$. Colors and numbers are arbitrary and specific to this figure. The correspondences are shown in (c), and will be used to construct the edges of the digraph $\mathcal D$ in LORA's random templex $\mathcal R$.}
    \label{fig:Tracking}
\end{figure*}

Figure \ref{fig:Tracking} shows how holes are tracked from one snapshot to the next using two successive snapshots of LORA, at $t=40.07$ and $t=40.075$. Tracking is performed by searching for the minimal distance between the barycenters of the minimal holes at consecutive snapshots. 
The symmetry of LORA is used in the tracking process. In Fig.~\ref{fig:Tracking}(a), $b_1$ and $b_2$ are symmetric at time $t=40.07$, and they must also be mapped to symmetric 1-holes $c_1$ and $c_2$ in the subsequent snapshot  at $t=40.075$,  Fig.~\ref{fig:Tracking}(b).

Applying this procedure to every hole in all the snapshots, we obtain the digraph of the random templex. The result for ten LORA snapshots in a time window $T_w=[40.065, 40.11]$ is shown in Fig.~\ref{fig:treeplot}. The digraph $\mathcal D$ here is not singly connected: $\cal{D}$ has fifteen connected components, and each of these directed subgraphs tells the story of one or several holes.  

Even if there are only ten snapshots within $T_w$, a single connected component in $\cal{D}$ may have more than ten nodes, because of the existence of merging and splitting events that enable connections between the storylines of different holes. Important LORA properties that can be extracted from its random templex are given in the next section. 

\section{Topological Tipping Points (TTPs)}
\label{sectemplex}

TTPs, introduced by \citet{charo2021noise}, can be now identified and classified using the digraph $\cal{D}$ of LORA's random templex. 

\begin{defi}
A {\bf Topological Tipping Point (TTP)} occurring at time $t^* \in [t_1,t_s]$ and at position $\bar{x}^*$ is encoded in the digraph $\cal{D}=(N,E)$ of a random templex $\cal{R}$ either (i) as a node that receives or emanates two or more edges or else (ii) as an initial or terminal node of a connected component in $\cal{D}$ that does not correspond to time $t_1$ or $t_s$ respectively.
\end{defi}

This definition allows one to  classify TTPs. The nodes that receive two or more edges are merging TTPs. The nodes that emanate more than one edge are splitting TTPs. Merging can be followed by splitting. The initial nodes of a connected component of the digraph that do not correspond to time $t_1$ are creation events. Destruction events are terminal nodes of a connected component that do not correspond to time $t_s$. 

\begin{figure*}
	\centering
	\includegraphics[width=0.7\textwidth]{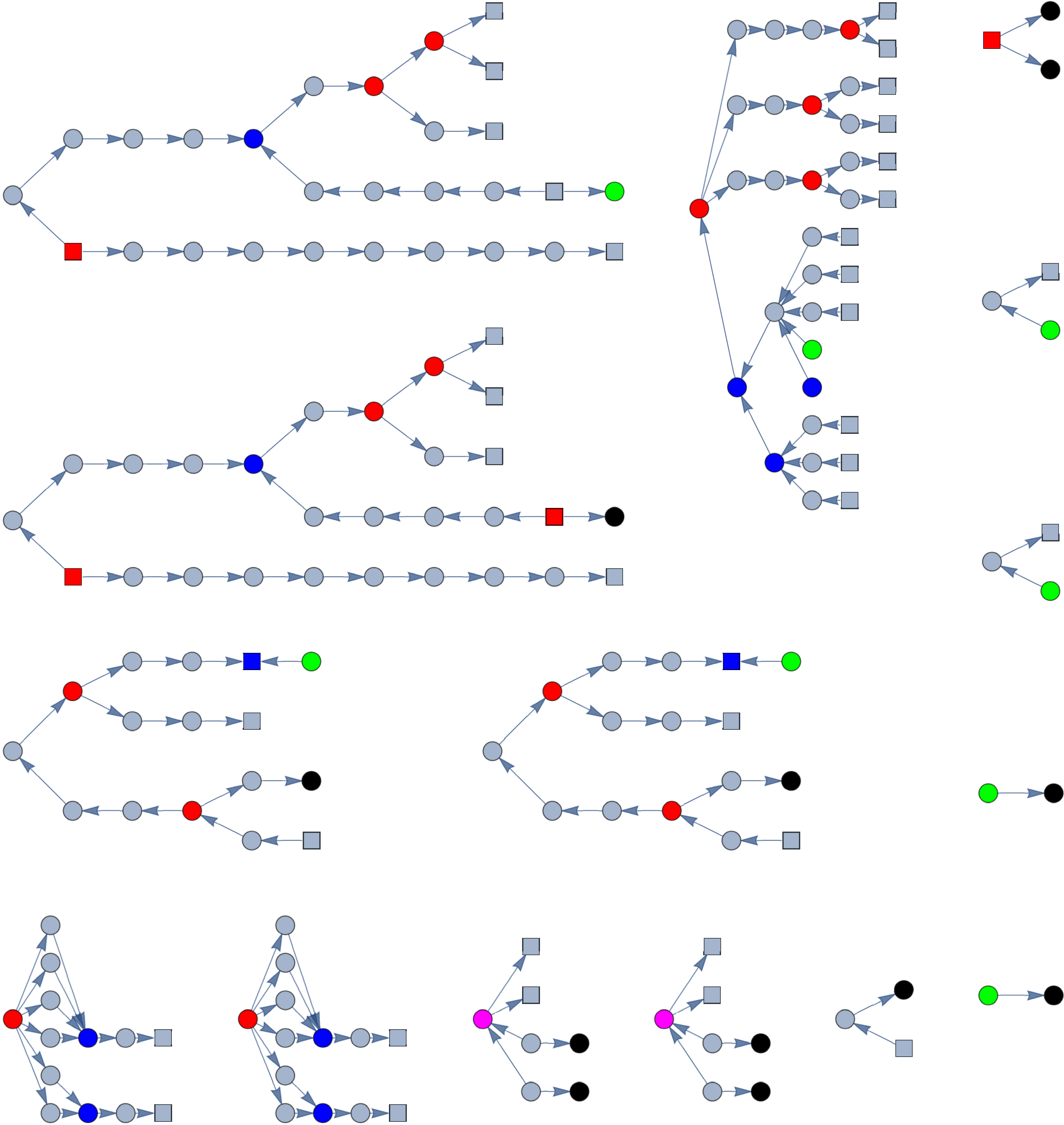}
	
	\caption{Tree plot of the digraph $\mathcal D$ of LORA's random templex $\cal{R}$ for $s=10$ snapshots in the time window $t_1 \le t \le t_s$. Square nodes correspond either to initial or final time nodes.  Tipping points can be identified and classified using $\cal{D}$. They are highlighted in different colors according to the type of event: creation in green, destruction in black, splitting in red, merging in blue, and merging followed immediately by splitting in magenta.  
}
	\label{fig:treeplot}
\end{figure*}

\begin{figure*}
\centering
\begin{subfigure}[b]{0.95\textwidth}
\includegraphics[width=0.6\textwidth]{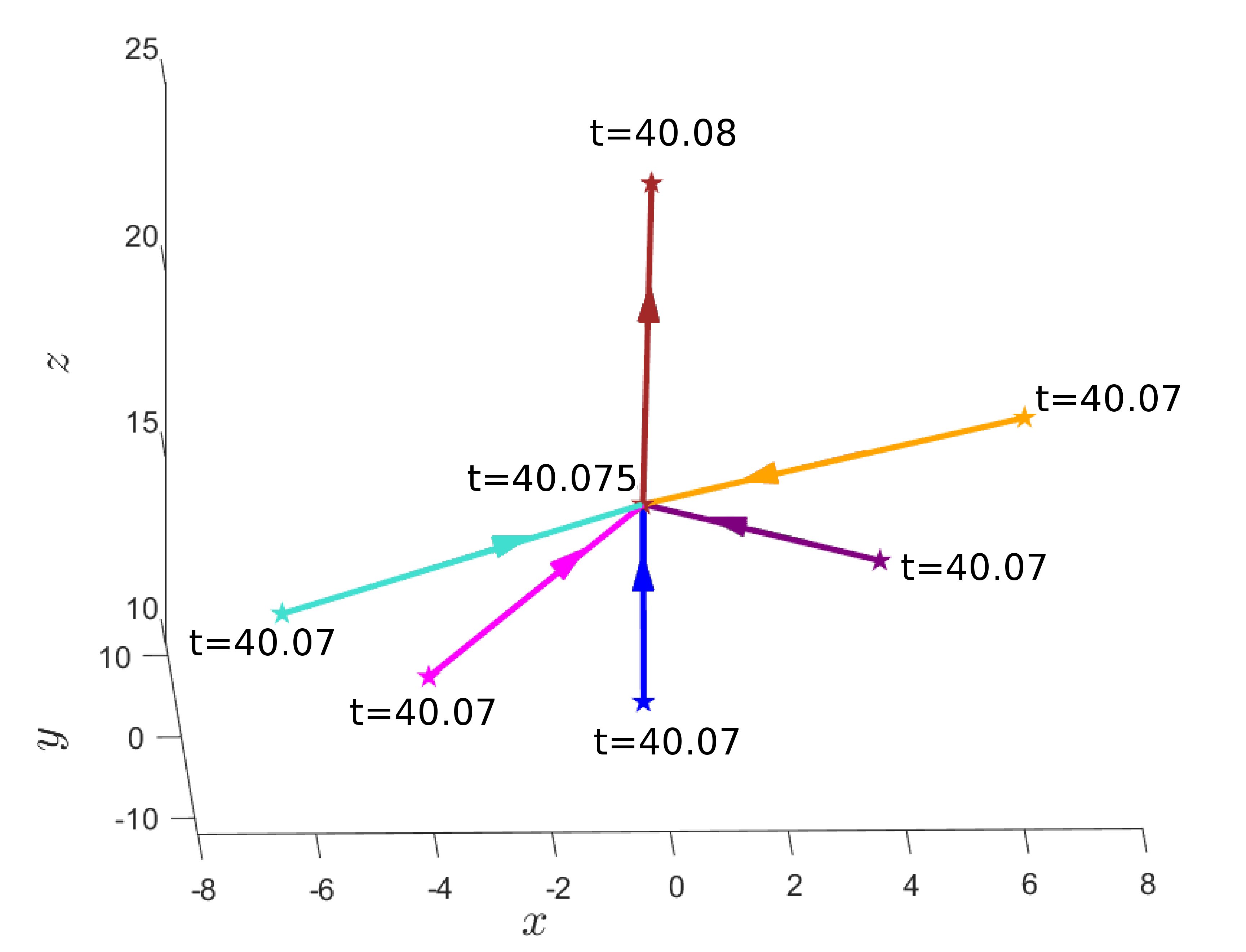}
 	\end{subfigure} \\
  \begin{subfigure}[b]{0.4\textwidth}
 \includegraphics[width=\textwidth]{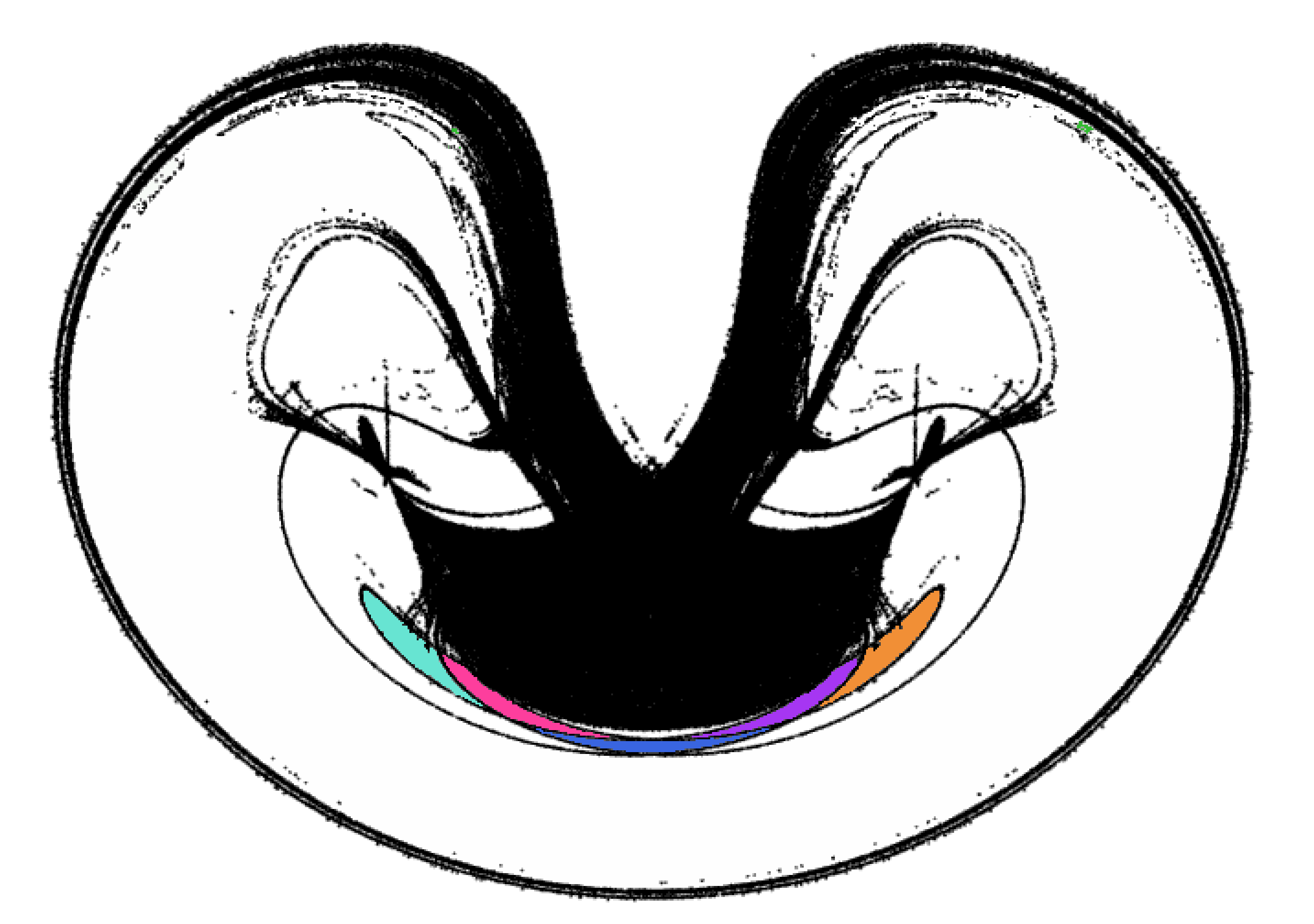}
 	\end{subfigure}
 \begin{subfigure}[b]{0.4\textwidth}
 	\includegraphics[width=\textwidth]{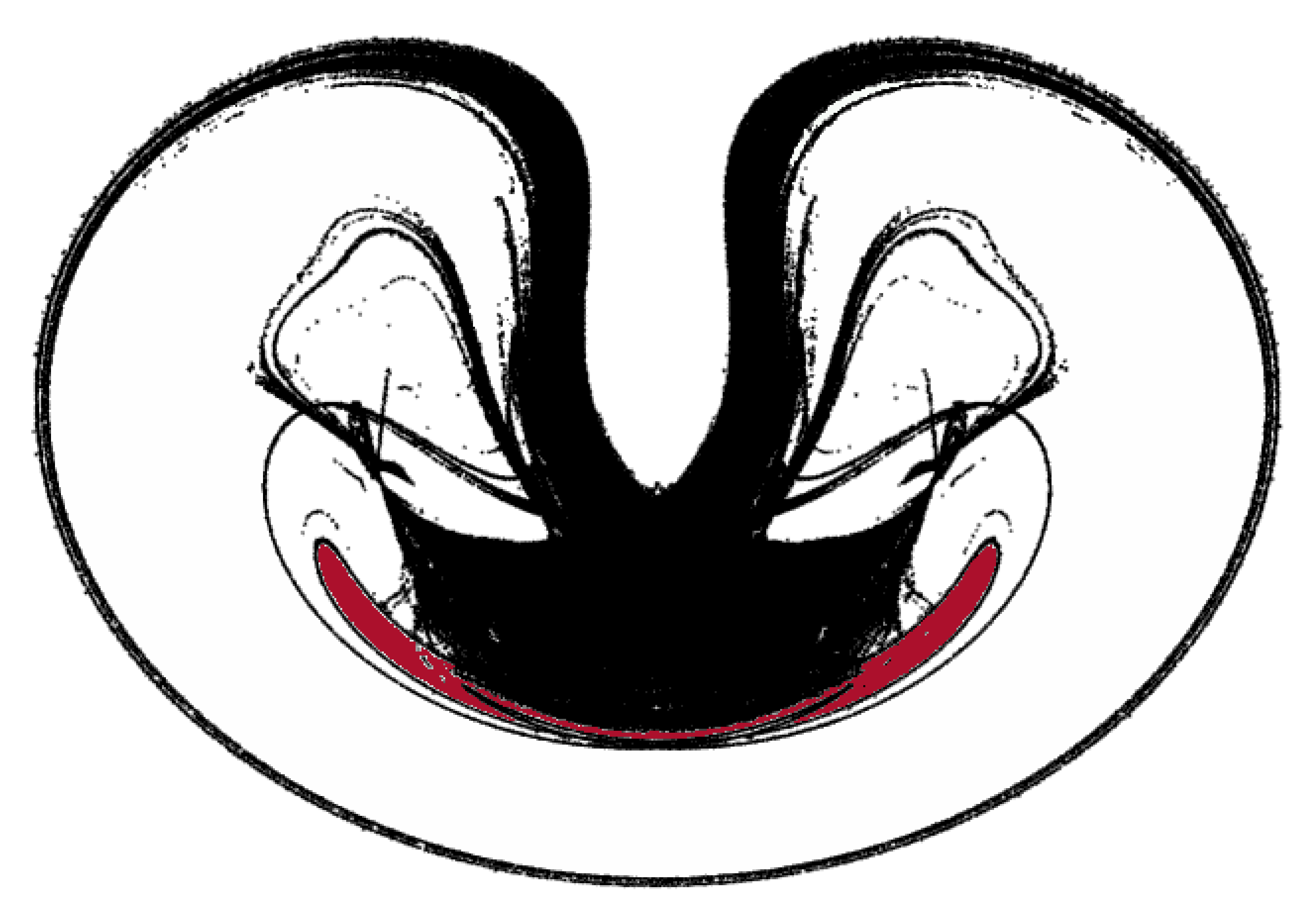}
 	\end{subfigure}
	\captionsetup{justification=justified}
   \caption{A merging event.  In the top panel, five holes that are distinct at $t=40.07$ are shown to merge into a single dark red hole at $t=40.075$. (Bottom panels) Phase portraits of the snapshots with the distinctly colored holes that merge into a single dark-red hole. 
   }
\label{fig:merging_node}
\end{figure*}

 \begin{figure*}
	
	\centering
	\begin{subfigure}[b]{0.95\textwidth}
		\includegraphics[width=0.6\textwidth]{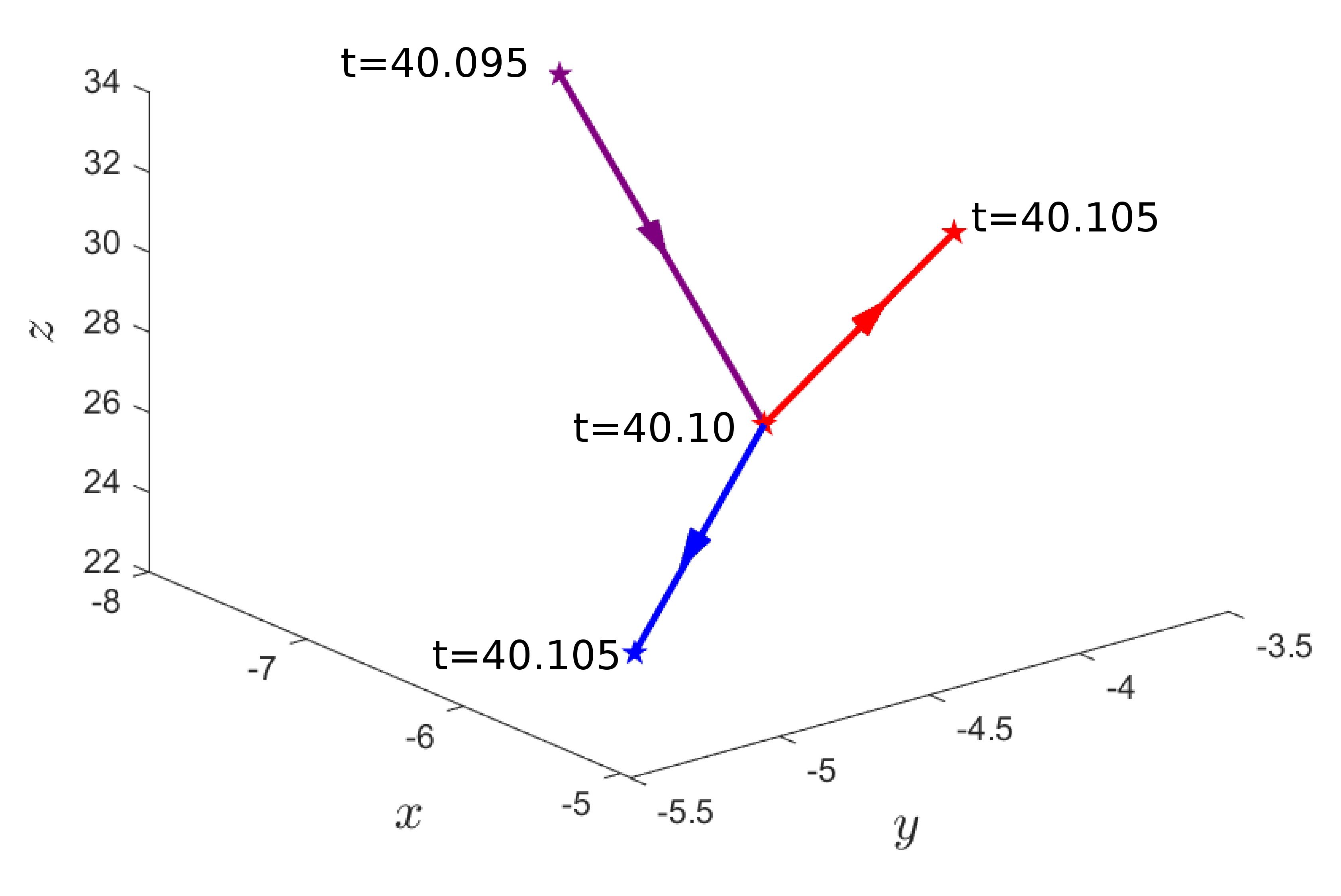}
	\end{subfigure} \\
	
	\begin{subfigure}[b]{0.4\textwidth}
		\includegraphics[width=\textwidth]{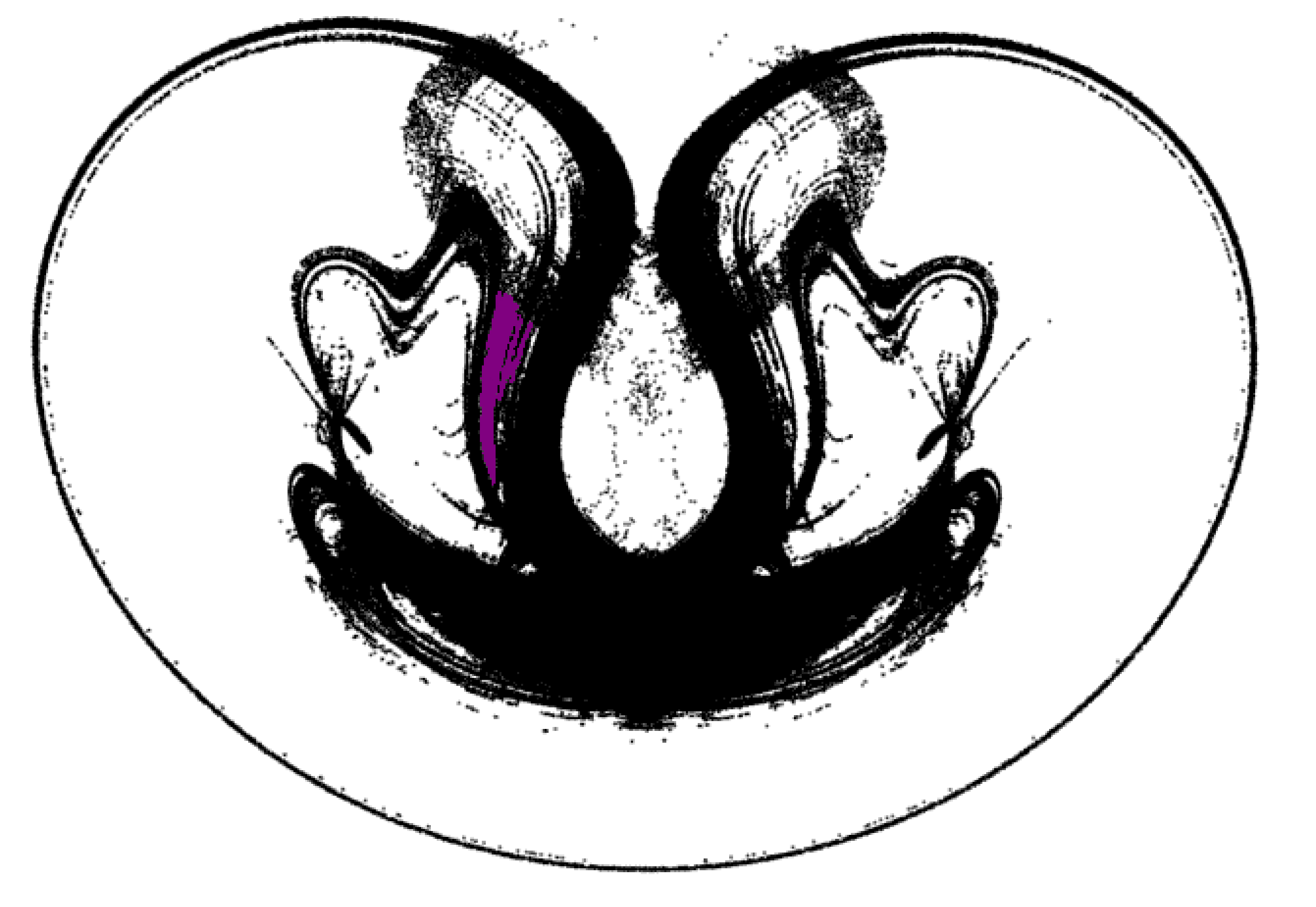}
	\end{subfigure}
	\begin{subfigure}[b]{0.4\textwidth}
		\includegraphics[width=\textwidth]{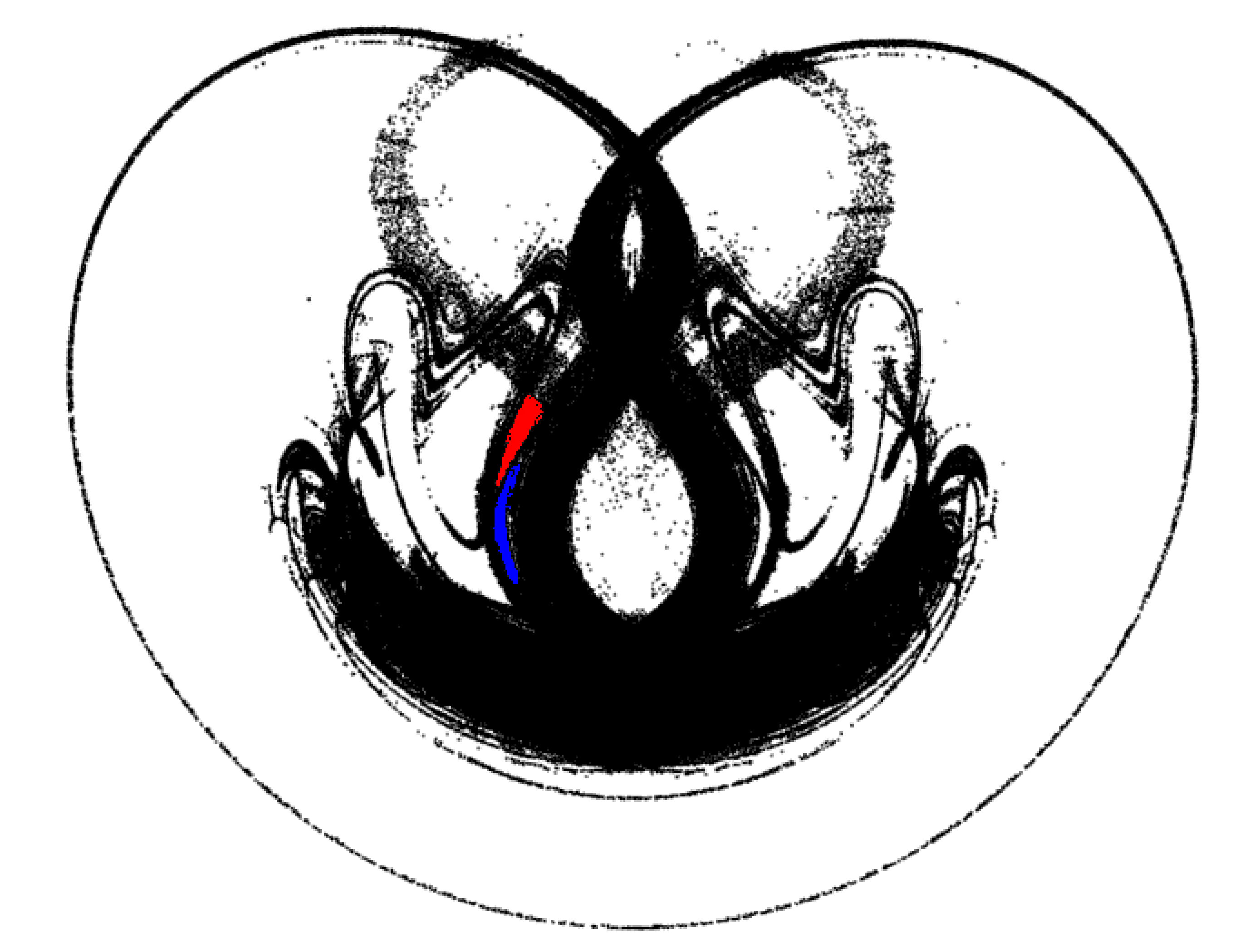}
	\end{subfigure}
	\captionsetup{justification=justified}
	\caption{A splitting event. (Top panel) a hole at $t=40.100$ splits into two different holes with different locations 
		in phase space at $t=40.105$. (Bottom panels) Phase portraits of the snapshots with the dark-red--colored 
		hole that splits into a blue and a fire red one.  
	}
	\label{fig:splitting_node}
\end{figure*}

Each tipping point is associated with a particular snapshot, that is a particular time instant in the time window $T_w$, and with a particular location in phase space, as defined by the barycenter of the corresponding hole. Figures \ref{fig:merging_node} and \ref{fig:splitting_node} show, in detail, an example of a merging node and of a splitting node, respectively. In both figures, the top plot shows the location of the nodes in phase space, as identified by the holes' barycenters, with an indication of the time instant, and the two bottom  plots show the phase portraits of the snapshots at which the holes merge or split being colored.  

The digraph of LORA's random templex $\cal{R}$ is shown in Fig.~\ref{fig:treeplot}. This tree corresponds to $s=10$ snapshots within $t_1 \le t \le t_s$ with $t_1=40.065$ and $t_s=40.11$.  Applying the definition of TTP to the nodes and edges in $\cal{D}$, one can detect them and classify them according to the type of event. We find 18 splitting TTPs (in red), 12 merging TTPs (in blue), 2 mergings followed by splittings (in magenta), 8 creation TTPs (in green), and 12 destruction TTPs (in black).

A better picture of how the holes are evolving in the system's phase space can be gained by using the coordinates of the barycenters of the holes for an embedding of $\mathcal D $ into this space. We call the embeddings of  $\mathcal D $'s distinct connected components into phase space {\bf constellations}. 

\begin{defi}
	A {\bf constellation} $\mathcal C$ is the set of immersed nodes and edges forming a connected component in the digraph $\mathcal D$ of a random templex. Each node is immersed in the phase space using the coordinates of the corresponding hole's barycenter. 
\end{defi}
\begin{figure*}
	\includegraphics[width=0.9\textwidth]{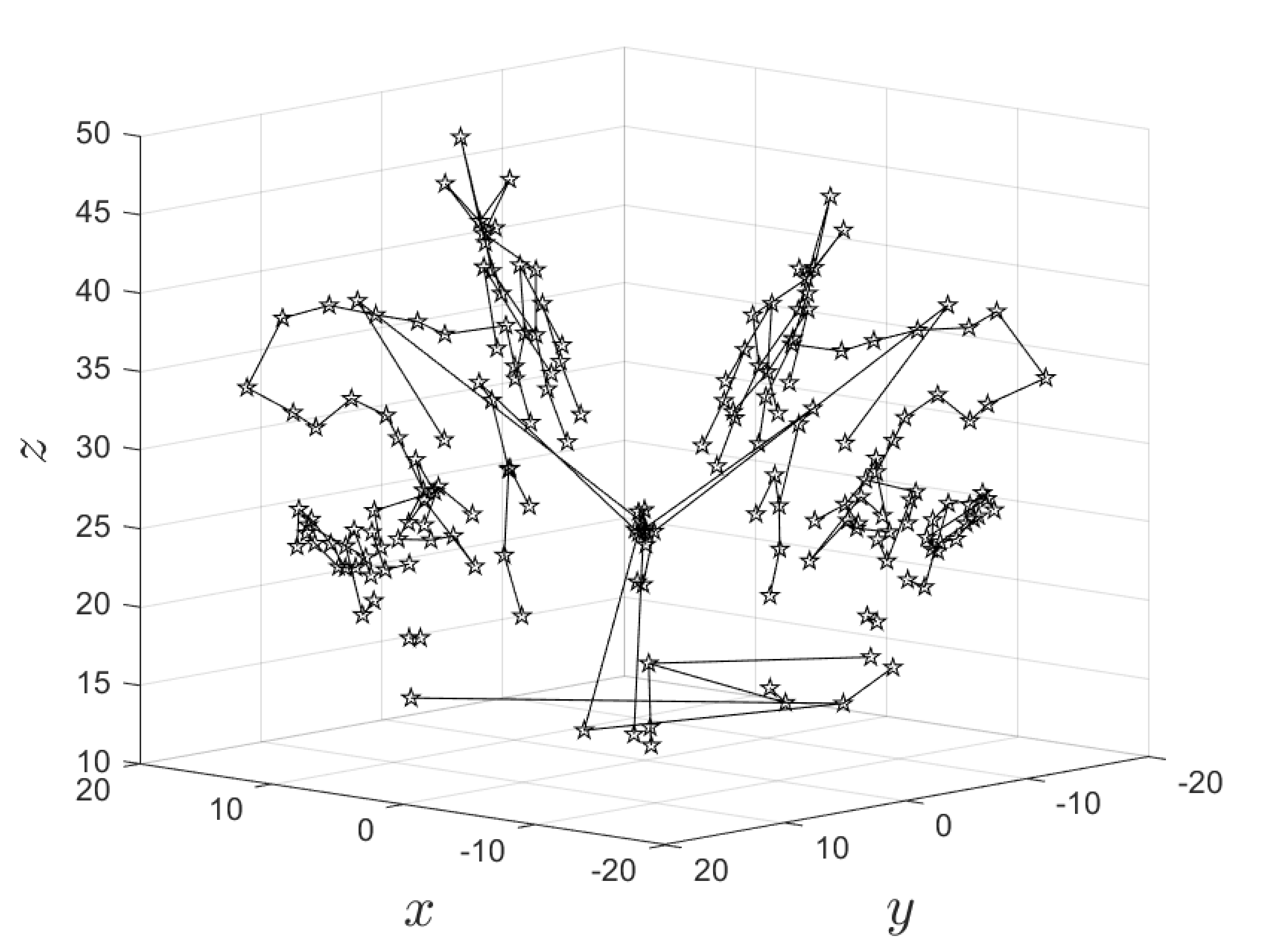}
\caption{The fifteen constellations that represent the essence of  the evolution of LORA's 1-holes within 
	the time window $T_w$. The plot shows the embedding of the digraph $\mathcal D$ of the random templex 
	$\mathcal R$ into the L63 model's phase space, by using the coordinates of the barycenters of the nodes $\mathcal N$.
}
\label{fig:constellations}
\end{figure*}

Figure~\ref{fig:constellations} presents the 15 constellations characterizing the evolution of LORA in $T_w$. 
While Fig.~\ref*{fig:treeplot} is a purely graph-theoretical representation of the evolution of LORA's random templex, Fig.~\ref{fig:constellations} is a step in connecting it with a more geometric representation of this evolution. Such a representation could provide a bridge between the random templex --- whose simplicity benefits from the invariance of the topology and of its breakdowns --- and a more detailed description of the flow's dynamics in phase space. 

\section{Concluding remarks} \label{sec:concl}

In this paper, we have introduced novel tools into algebraic topology 
and used them to provide insights into the behavior of random attractors, which combine deterministic chaos with stochastic perturbations \cite{crauel1994attractors,arnold1998random}. We summarize our results in the next subsection and discuss them in section~\ref{ssec:discuss}.

\subsection{Summary}
\label{ssec:summary}

The Introduction reviewed the basic tools of this trade: branched manifolds, cell complexes and homology groups \cite{kinsey2012topology} and it laid out the road plan for the paper. 

In section~\ref{sec2}, we presented the by now fairly well known theory for deterministic strange attractors with time-independent forcing. In section \ref{ssec:complex}, we illustrated the connection between the branched manifold that supports the Lorenz (L63) strange attractor's \cite{lorenz1963lorenz} invariant measure and a cell complex $K$, constructed by the Branched Manifold Analysis through Homologies (BraMAH) of \citet{sciamarella1999topological}; see Fig.~\ref{fig:det_lor}. We emphasized the independence of the homology groups ${\cal H}_k$ and associated Betti numbers $\beta_k$ from the details of the construction of the cell complex \cite{Poincare1895,kinsey2012topology,siersma2012poincare}. In particular, we showed that 1-generators, i.e., loops around a 1-hole with $\beta = 1$, needn't be minimal to still provide the homological information; see Fig.~\ref{fig:holes}. 

In section~\ref{ssec:templex}, we recalled the directed graph (digraph) $G$ associated with the cell complex $K$, as introduced by \citet{charo2022templex} for the L63 attractor, as well as for the spiral and funnel R\"ossler attractors, among others. This digraph $G$, with its nodes $N$ that are the cells and its edges $E$ that are the connections between them, is shown here for L63 in Fig.~\ref{fig:digraph_L63}; it provides the direction of the flow on the branched manifold from one cell to another. Together,  $K$ and $G$ form the templex $T=(K,G)$ associated with a particular nonlinear and, possibly, chaotic dynamics.

In section~\ref*{sec3}, we turned to our paper's main focus, namely extending the recent digraph and templex concepts and tools reviewed in section~\ref{sec2} to the study of random attractors that evolve in time \cite{crauel1994attractors,arnold1998random,romeiras1990multifractal}. To do so, we had to somehow incorporate the time into the definition of the digraph, by relating the cell complexes from one snapshot to another. This was implemented by constructing and labeling the minimal holes of one snapshot and tracking them to the next one; see Figs.~\ref{fig:MinCyc} and \ref{fig:Tracking}.

More precisely, the correspondence between snapshots was established by identifying the 1-generators of the homology groups, i.e., the 1-holes of two consecutive complexes. In order to implement hole tracking between snapshots, the minimal holes were defined using the algebraic procedure described in the appendix. These holes are the nodes $\mathcal N$ used to encode, through the edges $\mathcal E$ of the digraph $\mathcal D$, how the random attractor's invariant measure  evolves in time within a certain time window $T_w$. 

The random templex $\mathcal R$ is hence defined as the couple $\mathcal R = (\mathcal K, \mathcal D)$. The object $\mathcal K$ is the indexed family of cell complexes over the time interval $T_w$ under consideration, and the object $\mathcal D$ is the tree illustrated in Fig.~\ref{fig:treeplot}, with nodes $\mathcal N$ at each snapshot and edges $\mathcal E$ connecting one snapshot with another.

Each node of the digraph is a minimal hole at a certain time. Each hole has the possibility of connecting with itself in the subsequent snapshot, but also with other holes, through mergings or splittings. Holes can disappear as well from a snapshot to the next, or be created at an intermediate snapshot. Each connected component of the digraph $\mathcal D$ tells the story of how a hole evolves and how it connects to other holes, as time progresses; see Fig.~\ref{fig:treeplot}. Examples of merging and splitting of minimal 1-holes were given in Figs.~\ref{fig:merging_node} and \ref{fig:splitting_node}, respectively. 

Much of the motivation of the work herein had to do with the visually striking, rapid changes in the evolution of LORA's invariant measure as described by \citet{chekroun2011stochastic}. These visually rapid changes in time were documented by changes in LORA's algebraic topology; see Fig.~4 in \citet{charo2021noise}, in which $\mathcal H_1$ changed from $\mathbb{Z}^2$ to $\mathbb{Z}^{10}$ and back to $\mathbb{Z}^4$ in very small time steps of $\Delta t = 0.09$. 

We thus addressed in section~\ref{sectemplex}, using the more sophisticated topological apparatus of section~\ref{sec3}, the existence and nature of topological tipping points (TTPs) conjectured by \citet{charo2021noise}. A TTP occurring at time $t^*$ and at position $\bar{x}^*$ is encoded in the digraph $\cal{D}=(N,E)$ of a random templex $\cal{R}$ either (i) as a node that receives or emanates two or more edges or else (ii) as an initial or terminal node of a connected component in $\cal{D}$ that does not correspond to the beginning or end of the time interval $T_w$ under consideration. 

First of all, by considering time steps of $\Delta t = 0.005$ that are even smaller compared to the L63 model's \cite{lorenz1963lorenz} characteristic time of order unity, we confirmed, at least mumerically, than the holes actually appear, disappear and merge or split instantaneouly. While a truly rigorous mathematical proof that this is indeed so might be difficult, the numerical evidence is rather overwhelming. Another matter that is left open at this point is whether changes in torsion --- such as between a M\"obius band, with torsion, and a regular one, without it --- might be as sudden or not.

\vspace{-0.1cm}

\subsection{Discussion}
\label{ssec:discuss}

Given the fairly novel and even surprising nature of the concepts and tools introduced herein, a number of other questions are worth mentioning. One concerns the possibility of establishing closer connections between the metric flow of the solutions of a dynamical system in phase space and the topological description provided herein. The constellations shown in Fig.~\ref{fig:constellations} and discussed at the end of section~\ref{sectemplex} suggest such a possibility, since one uses the embedding of the digraph $\mathcal D$ of the random templex $\mathcal R$ into the L63 model's phase space, by using the coordinates of the barycenters of the nodes $\mathcal N$. 

This representation emphasizes how a  random templex contains the information of when --- i.e., in which snapshot --- and where i.e., in which phase space location --- topological tipping is taking place. \citet{charo2021noise} emphasized already that BraMAH-based topological data analysis \cite{sciamarella1999topological,sciamarella2001unveiling} is not restricted to low-dimensional dynamical systems. Clearly, some forms of reduced-order modeling \cite{kondrashov2015data, santos2021reduced} will be necessary to treat models with substantial geographical resolution. But it is at least conceivable to use the mixed localization approach of our constellations for the study of tipping in subsystems of a larger system, in the spirit of the rather popular “tipping elements” of \citet{Lenton.ea.2008}.

\vspace{-0.5cm}

\section{Acknowledgments} 

\vspace{-0.3cm}

It is a pleasure to acknowledge stimulating discussions with M.D. Chekroun on extending the results of \citet{charo2021noise} to an improved detection of topological tipping points.
This work is supported by the French National program LEFE (Les Enveloppes Fluides et l’Environnement) and by the CLIMAT-AMSUD 21-CLIMAT-05 project (D.S.). G.D.C. gratefully acknowledges her postdoctoral scholarship from CONICET. The present paper is TiPES contribution \# xy; this project has received funding from the EU Horizon 2020 research and innovation program under grant agreement No. 820970, and it helps support the work of M.G.

\vspace{-0.5cm}

\section*{Data availability}
The data that support the findings of this study are available within the article and its supplementary material. The code that computes templex properties is available at
\href{https://git.cima.fcen.uba.ar/sciamarella/templex-properties.git}{git.cima.fcen.uba.ar/sciamarella/templex-properties.git}.

\appendix
\section{Computing homologies and stripexes from a templex}
\label{homologies}

\begin{figure}
	\includegraphics[width=0.8\linewidth]{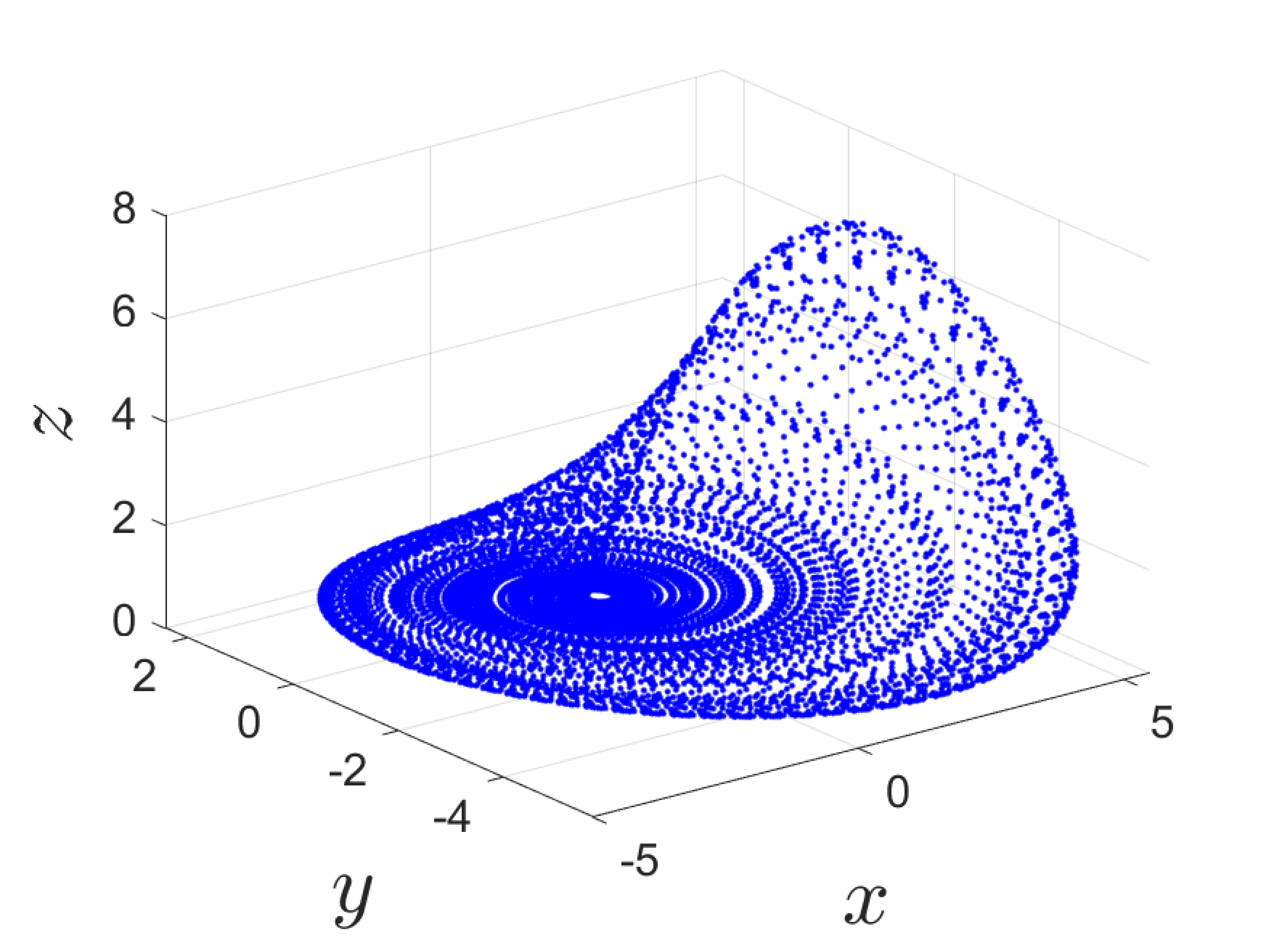}
	\caption{Chaotic attractor of the R\"ossler system \eqref{roseq76}. The parameter values are $a=0.43295$, $b = 2$, and $c=4$.
	}
	\label{fig:rosslerattractor}
\end{figure}

This appendix details the algebraic computations that enable one to extract the topological features --- such as homologies, joining loci, and stripexes --- from a templex. For the sake of simplicity, we consider the chaotic attractor produced 
by the R\"ossler system \cite{Ros76c} with  $a=0.43295$, $b = 2$ and $c=4$. 
\begin{equation}
  \label{roseq76}
  \left\{
    \begin{array}{l}
      \dot{x} = -y -z \\[0.1cm]
      \dot{y} = x + ay \\[0.1cm]
	    \dot{z} = b + z (x-c)  \, .
    \end{array}
  \right.
\end{equation}
A point cloud of this spiral attractor appears in Fig.~\ref{fig:rosslerattractor}.

\begin{figure}
\centering
\begin{subfigure}[b]{0.4\textwidth}
 \includegraphics[width=\linewidth]{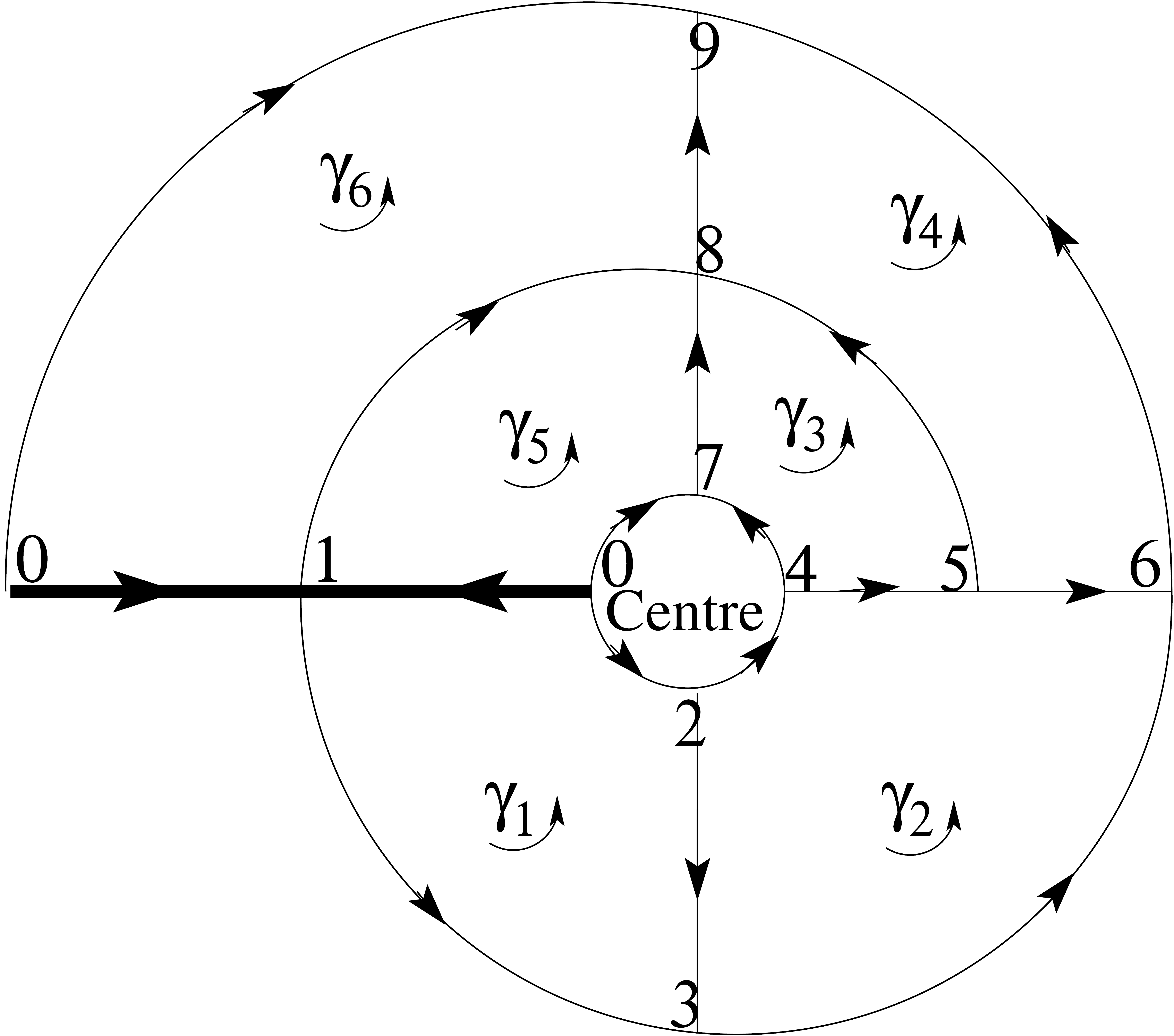}
 \caption*{(a)}
\end{subfigure} 

\begin{subfigure}[b]{0.25\textwidth}
\includegraphics[width=\linewidth]{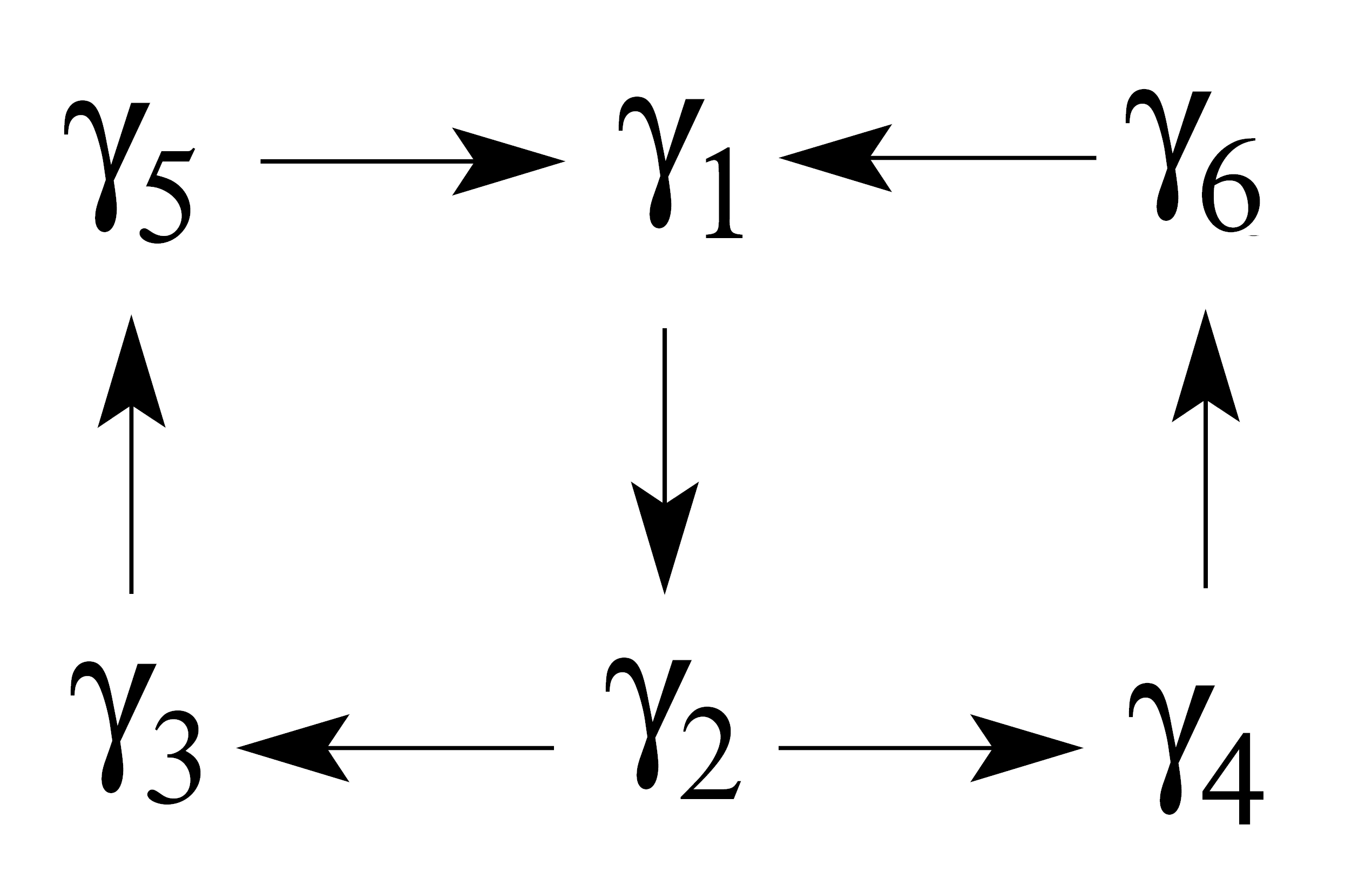}
 \caption*{(b)}
\end{subfigure} 

 \caption{A templex $T'=(K',G')$ for the R\"ossler attractor. 
    (a) Planar diagram of the oriented complex $K'$.
    (b) Digraph $G'$ indicating the connections between the 2-cells in $K'$. }
	\label{fig:rosslertemplex}
\end{figure}

A cell complex $K'$ for this system is shown in Fig.~\ref{fig:rosslertemplex}(a) using a planar diagram of the complex, i.e., a diagram in which the cells that appear to be duplicated are in fact two copies of the same cell and must therefore be glued together and considered as one. The advantage of such a planar diagram over the attractor's three-dimensional representation is that one can see the whole structure, whereas if we plot the complex in three dimensions --- juxtaposed on the point cloud as in Fig.~\ref{fig:det_lor} --- some parts of the plot are obscured by the perspective. In the case of $K'$, the $1$-cells that must be glued together are drawn with heavy lines. If this planar diagram is drawn on paper, one can obtain a model of the attractor by gluing the heavy lines together. 

\begin{equation}
  M_1^T =
   \begin{array}{crrrrrrrrrr}
    \hline \hline
	  \\[-0.3cm]
    \partial_1 &\langle 0  \rangle& \langle 1  \rangle&\langle 2  \rangle &\langle 3  \rangle&\langle 4  \rangle&\langle 5 \rangle &\langle 6  \rangle&\langle 7  \rangle& \langle 8 \rangle & \langle 9 \rangle
	  \\[0.1cm]
    \hline
	  \\[-0.3cm]
        \langle 0,1 \rangle & -1 & 1 & 0 & 0&0 &0 &0 & 0&0 &0\\
	  \langle 0,2 \rangle & -1 & 0 &1 & 0&0 &0 &0 &0 &0 &0\\
	  \langle 0,7 \rangle & -1 & 0 & 0&0 &0 &0 &0 &1 &0 &0\\
	    \langle 0,9 \rangle & -1 & 0 &0 & 0&0 &0 &0 &0 & 0&1\\
	    \langle 1,3 \rangle & 0 &-1 &  0& 1& 0&0& 0& 0& 0&0\\
	  \langle 1,8 \rangle & 0&-1 &0 & 0& 0& 0& 0&0 & 1&0\\
	  \langle 2,3 \rangle & 0& 0& -1&1 & 0 &0 & 0& 0& 0&0\\
	  \langle 2,4 \rangle &0 & 0 & -1&0 & 1& 0& 0& 0& 0&0\\
	  \langle 3,6 \rangle & 0& 0& 0 &-1 &0 & 0& 1 &0 & 0&0\\
	  \langle 4,5 \rangle &0 &0 &0 &0 &-1 &1 &0 & 0 & 0&0\\
	  \langle 4,7 \rangle & 0& 0& 0& 0& -1 &0 & 0& 1&0 &0\\
	  \langle 5,6 \rangle &0 &0 &0 & 0& 0 & -1 &1 &0 &0 &0\\
	  \langle 5,8 \rangle &0 &0 & 0& 0& 0& -1&0 & 0& 1&0\\
	  \langle 6,9 \rangle & 0&0 & 0& 0& 0& 0 &-1 & 0& 0&1\\
         \langle 7,8 \rangle & 0& 0&0 &0 &0 & 0& 0& -1&1 &0\\
	  \langle 8,9 \rangle &0 &0 &0 & 0&0 &0 & 0 & 0 &-1 &1\\[0.1cm]
    \hline \hline
  \end{array}
  \label{border1}
\end{equation}

\begin{equation}
  M_2 =
  \begin{array}{crrrrrr}
    \hline \hline
	  \\[-0.3cm]
    \partial_2 & \gamma_1 & \gamma_2 & \gamma_3 & \gamma_4  & \gamma_5 & \gamma_6
	  \\[0.1cm]
    \hline
	  \\[-0.3cm]
	  \langle 0,1 \rangle &  1 & 0 & 0  & 0 & -1 & 1 \\
	  \langle 0,2 \rangle & -1 & 0 &  0 & 0 & 0& 0\\
	  \langle 0,7 \rangle & 0 &  0 &  0 & 0 & 1& 0\\
	  \langle 0,9 \rangle &  0 & 0 & 0 & 0 & 0& -1\\
	  \langle 1,3 \rangle &  1 & 0 & 0 & 0 & 0& 0\\
	  \langle 1,8 \rangle &  0 & 0 & 0 & 0 & -1& 1\\
	  \langle 2,3 \rangle & -1 & 1 & 0 & 0& 0& 0\\
	  \langle 2,4 \rangle & 0 & -1 & 0 & 0& 0&0\\
	  \langle 3,6 \rangle & 0 & 1 & 0 & 0 & 0&0\\
	  \langle 4,5 \rangle & 0 & -1 & 1 & 0& 0&0\\
	  \langle 4,7 \rangle & 0 & 0 & -1 & 0& 0&0\\
	  \langle 5,6 \rangle & 0 & -1& 0 & 1 & 0&0\\
	  \langle 5,8 \rangle & 0 & 0 & 1 & -1& 0&0\\
        \langle 6,9 \rangle & 0 & 0 & 0 & 1 & 0&0\\
        \langle 7,8 \rangle & 0 & 0 & -1& 0 & 1&0\\
        \langle 8,9 \rangle & 0 & 0 & 0 & -1& 0&1\\[0.1cm]
    \hline \hline
  \end{array}
  \label{border2}
\end{equation}

The complex $K'$ is made up of six 2-cells: 
\begin{align*}
\gamma_1=<1, 3, 2, 0>,   \gamma_2 = <2, 3, 6, 5, 4>, \gamma_3 = <4, 5, 8, 7>,\\
 \gamma_4 = <5, 6, 9, 8>,\gamma_5 =~<7, 8, 1, 0>, \gamma_6=<1, 8, 9, 0>.
\end{align*} 
 The 2-cell denoted by $\gamma_6$ is attached through
the 1-cell $<0,1>$ to the cells $\gamma_1$ and $\gamma_5$ with a gluing direction that highlights the folding that is taking place.

 The boundary operator $ \partial_k$ computes the boundary of an oriented k-cell
$ \partial_k : {\cal C}_k\rightarrow {\cal C}_{k-1} \,$, with $C_k = \sum_i a_i \, \sigma^i_k$, where $a_i \in \mathbb{Z}$ and $\sigma^i_k$ is a $k$-cell indexed by
$i \in \mathbb{N}$. The action of the boundary maps is the following: 
\begin{align*}
\partial_0 <1>&= 0, \quad \partial_1<0,1>= <1> - <0>,\quad\\
\partial_2 <1, 8, 9, 0>&= <1,8> + <8,9> - <0,9> + <0,1>.
\end{align*} 
The information about the boundary of all the cells can  be condensed within two matrices: $M_1$ in Eq.~\eqref{border1} represents the boundary of the 1-cells, i.e., it relates the 1-cells with the 0-cells, while $M_2$ in Eq.~\eqref{border2}  represents the boundary of the 2-cells that relates the 2-cells with the 1-cells. 

Using these matrices, it  is possible to compute the quotient group
\[
  {\cal H}_k  := {\cal Z}_k / {\cal B}_k  \, , 
\]
called the $k$-th homology group of a complex $K$.
 For instance, ${\cal H}_1 := {\cal Z}_1 / {\cal B}_1$, where
$B_1$ can be computed as the linearly independent rows of the  transpose of $M_2$, and $Z_1$, the set of  1-cycles, is the null space of the transpose of $M_1$. If the elements of $B_1$ are expressed as linear combinations of the elements of $Z_1$, the result is the list of chains of 1-cycles that are homologous to zero. This yields the homology relations between the 1-cycles, which lead, in turn, to the 1-generators, namely the generators of $H_1$. 

In the case of $K'$, there is only one 1-generator:
$$h_1=+<0, 2>+<2, 4>+<4, 7>-<0, 7>.$$ 
Notice that this 1-hole is minimal, since it contours the hole corresponding to the focus-type hole in the attractor. In short, all 1-cycles in $K'$ are homologous to $h_1$. For $K'$, the lower and higher levels, $H_0$ and $H_2$, equal the empty set, since the complex has only one connected component and no cavities. Furthermore, in this complex, like in $K$ for L63, there are no orientability chains and therefore no torsion group. 

Up to this point, we have calculated homologies as customary in any algebraic topology textbook \cite{kinsey2012topology}. But the cell complex contains information relevant to the characterization of a chaotic attractor that is neither included in the homology groups nor in the torsion groups per se.
Locating the $1$-cells which are shared by at least three $2$-cells, we can also compute the joining 1-cells. In $K'$, $<0,1>$ is such a joining 1-cell. The $2$-cells that are glued to it can also be identified: they are $\gamma_1$, $\gamma_5$ and $\gamma_6$. Together they form what we call the joining $2$-chain:
$\gamma_1-\gamma_5+\gamma_6=<1,3,2,0>-<7,8,1,0>+<1,8,9,0>$. 
 
Let us now endow the complex $K'$ shown in Fig.~\ref{fig:rosslerattractor} (a) with a directed graph $G'$, shown in Fig.~\ref{fig:rosslerattractor} (b), in order to form the templex $T'$ of the R\"ossler attractor in Fig.~\ref{fig:rosslerattractor}. The set of operations required to combine the complex $K'$ and the digraph $G'$ are detailed in \citet{charo2022templex}. The joining 1-cell has two ingoing 2-cells, $\gamma_5$ and $\gamma_6$, and one outgoing 2-cell, $\gamma_1$. This information can be viewed in  the joining subgraph $\{\gamma_5 \rightarrow \gamma_1, \gamma_6 \rightarrow \gamma_1\}$ of Fig.~\ref{fig:rosslertemplex}(b).

The existence of an orientation change within a connected set of joining 1-cells reveals the existence of a  splitting 0-cell. In this example, no splitting 0-cells are found for $T'$. The stripexes are given by:
\begin{subequations}
	\begin{align}
& \gamma_1 \rightarrow \gamma_2 \rightarrow \gamma_4 \rightarrow \gamma_6 
  \rightarrow \gamma_1, \label{stripex1R} \\
& \gamma_1 \rightarrow \gamma_2 \rightarrow \gamma_3 \rightarrow \gamma_5  \rightarrow \gamma_1.  \label{stripex2R} 
  	\end{align}
\end{subequations}

The R\"ossler attractor is thus homologically equivalent to a cylinder, but a cylinder as such does not have a joining locus. The templex $T'$ tells us much more about the attractor's structure: its joining locus has a single component --- as expected for an  attractor bounded by a genus-1 torus --- and there are two stripexes, $\mathcal{S}_1$ given by \eqref{stripex1R} and ${\cal S}_2$ given by \eqref{stripex2R}. 
The stripex ${\cal S}_1$ is a M\"obius band while $S_2$ is a cylinder or normal band (without 
torsion). 

All the computations herein can be handled algorithmically with the Wolfram Mathematica code provided as supplementary material, and freely available at \href{https://git.cima.fcen.uba.ar/sciamarella/templex-properties.git}{git.cima.fcen.uba.ar/sciamarella/templex-properties.git}.\\ 

\section*{References}
\bibliography{LORA_topol-MG_v1}

\end{document}